\documentclass[a4paper,fleqn,usenatbib]{mnras}

\usepackage{newtxtext,newtxmath}

\usepackage[T1]{fontenc}
\usepackage{ae,aecompl}

\usepackage{graphicx}	
\usepackage{amsmath}	
\usepackage{amssymb}	
\usepackage{CJKutf8}	

\newcommand{\hst}{{\it HST\/}}       
\newcommand{\herschel}{{\it Herschel\/}}       
\newcommand{\chandra}{{\it Chandra\/}}

\newcommand{\jwst}{{\it JWST\/}}
\newcommand{\spitzer}{{\it Spitzer\/}}

\newcommand{\xray}{\hbox{X-ray}}  
\newcommand{\cdfs}{\hbox{CDF-S}}

\newcommand{\lx}{L_{\rm X}}

\newcommand{\kbol}{k_{\rm bol}}

\newcommand{\MsunYr}{M_\odot\ \mathrm{yr^{-1}}}
\newcommand{\lbol}{L_{\rm bol}}

\newcommand{\mbh}{M_{\rm BH}}

\newcommand{\bharbar}{\overline{\rm BHAR}}
\newcommand{\sfrbar}{\overline{\rm SFR}}
\newcommand{\mstar}{M_{\star}}
\newcommand{\mlim}{M_{\rm lim}}
\newcommand{\mbulge}{M_{\rm bulge}}

\title[SMBH-bulge coevolution]{Evident black hole-bulge coevolution in the distant universe}

\author[G. Yang et al.]{
G. Yang \begin{CJK*}{UTF8}{gbsn}(杨光)\end{CJK*},$^{1,2}$\thanks{E-mail: gyang206265@gmail.com (GY)}  
W. N. Brandt,$^{1,2,3}$
D. M. Alexander,$^{4}$
C.-T. J. Chen \begin{CJK*}{UTF8}{bsmi}(陳建廷)\end{CJK*},$^{5}$\newauthor
Q. Ni \begin{CJK*}{UTF8}{gbsn}(倪清泠),\end{CJK*}$^{1,2}$
F. Vito,$^{6}$
and
F.-F. Zhu \begin{CJK*}{UTF8}{gbsn}(朱飞凡)\end{CJK*}$^{1,7}$
\\
$^{1}$Department of Astronomy and Astrophysics, 525 Davey Lab, The Pennsylvania State University, University Park, PA 16802, USA\\
$^{2}$Institute for Gravitation and the Cosmos, The Pennsylvania State University, University Park, PA 16802, USA\\
$^{3}$Department of Physics, 104 Davey Laboratory, The Pennsylvania State University, University Park, PA 16802, USA\\
$^{4}$Centre for Extragalactic Astronomy, Department of Physics, Durham University, South Road, Durham DH1 3LE, UK\\
$^{5}$Marshall Space Flight Center, Huntsville, AL 35811, USA\\
$^{6}$Instituto de Astrof{\'{\i}}sica and Centro de Astroingenier{\'{\i}}a, Facultad de F{\'{i}}sica, Pontificia Universidad Cat{\'{o}}lica de Chile, Casilla 306, Santiago 22, Chile\\
$^{7}$CAS Key Laboratory for Research in Galaxies and Cosmology,
Center for Astrophysics, Department of Astronomy,\\ 
University of Science and Technology of China, Chinese Academy of 
Sciences, Hefei, Anhui 230026, China
}

\date{Accepted XXX. Received YYY; in original form ZZZ}

\pubyear{2018}

\begin{document}
\label{firstpage}
\pagerange{\pageref{firstpage}--\pageref{lastpage}}
\maketitle

\begin{abstract}
Observations in the local universe show a tight correlation between
the masses of supermassive black holes (SMBHs; $\mbh$) and host-galaxy 
bulges ($\mbulge$), suggesting a strong connection between SMBH and bulge growth.
However, direct evidence for such a connection in the distant universe 
remains elusive.
We have studied sample-averaged SMBH accretion rate ($\bharbar$)
for bulge-dominated galaxies at $z=0.5\text{--}3$.
While previous observations found $\bharbar$ is strongly related
to host-galaxy stellar mass ($\mstar$) for the overall galaxy population, 
our analyses show that, for the bulge-dominated population, $\bharbar$ 
is mainly related to SFR rather than $\mstar$.  
{This ${\bharbar}$-SFR relation is highly significant, e.g.\ 
${9.0\sigma}$ (Pearson statistic) at ${z=0.5\text{--}1.5}$.}
Such a $\bharbar$-SFR connection does not exist among 
our comparison sample of galaxies that are not bulge-dominated, 
for which $\mstar$ appears to be the main determinant of SMBH accretion. 
This difference between the bulge-dominated and comparison
samples indicates that SMBHs only coevolve with bulges rather
than the entire galaxies, 
{explaining the tightness of the local 
${\mbh}$-${\mbulge}$ correlation.
Our best-fit ${\bharbar}$-SFR relation for the bulge-dominated 
sample is ${\log\bharbar = \log\mathrm{SFR} - (2.48\pm0.05)}$
(solar units).}
The best-fit $\bharbar/\mathrm{SFR}$ ratio (${10^{-2.48}}$) for 
bulge-dominated galaxies is similar to the observed $\mbh/\mbulge$ values in the 
local universe.
Our results reveal that SMBH and bulge growth are in lockstep,
and thus non-causal scenarios of merger averaging are unlikely
the origin of the $\mbh$-$\mbulge$ correlation.
This lockstep growth also predicts that the $\mbh$-$\mbulge$ relation
should not have strong redshift dependence.
\end{abstract}



\section{Introduction}
\label{sec:intro}
One of the essential challenges of extragalactic astronomy is to understand
the connection between supermassive black holes (SMBHs) and their 
host galaxies. 
It is well established that the masses of SMBHs ($\mbh$) are tightly 
correlated with the stellar masses of host-galaxy classical bulges 
($\mbulge$) in the local universe (the Magorrian relation; e.g. 
\hbox{\citealt{magorrian98}}; \hbox{\citealt{ferrarese00}}; 
\hbox{\citealt{gebhardt00}}; \hbox{\citealt{haring04}};
\hbox{\citealt{kormendy13}}).
The intrinsic scatter of the $\mbh$-$\mbulge$ correlation is only 
$\approx 0.3$~dex \citep{kormendy13}. 
This tight correlation is surprising considering that $\mbh$ is only
a tiny fraction (a few thousandths) of $\mbulge$. 
Therefore, some fundamental connections between the growth of 
SMBHs and host-galaxy bulges likely exist over cosmic history.  
These physical connections are often termed as ``SMBH-bulge 
coevolution''.

Tremendous observational efforts have been made to identify these 
mysterious connections. 
It has been found that the cosmic evolution of SMBH accretion
rate (BHAR) density and star formation rate (SFR) density are broadly
similar, both peaking at $z\sim 2$ \citep[e.g.][]{aird10, aird15,
mullaney12, madau14, ueda14}.
However, from observations of active galactic nuclei (AGNs), 
the SFR is a relatively flat function of the observed BHAR at a 
given redshift \citep[e.g.][]{harrison12, rosario13b, azadi15, 
stanley15, lanzuisi17, stanley17, dai18}.
This apparent lack of a strong SFR-BHAR connection might be 
caused by AGN variability.
While star formation activity is stable on time scales longer
than $\sim 100$~Myr, SMBH accretion could vary strongly on much
shorter time scales ($\approx 10^2\text{--}10^7$~yr; 
e.g. \hbox{\citealt{martini04}}; \hbox{\citealt{kelly10}};
\hbox{\citealt{novak11}}; \hbox{\citealt{sartori18}}).
An intrinsic connection between SFR and long-term 
average BHAR might be hidden by this strong AGN variability.

To obtain long-term average BHAR, the ideal way is to observe
a galaxy for at least millions of years, which is presently
infeasible.
Practically, it has been proposed to adopt sample-averaged BHAR 
($\bharbar$) as a proxy of long-term average BHAR 
\citep[e.g.][]{chen13, hickox14}. 
Indeed, a positive $\bharbar$-SFR connection has been observed
\citep[e.g.][]{chen13, hickox14, lanzuisi17, yang17}.
However, \hbox{\citet{yang17}} show, via partial correlation 
analyses (PCOR), that $\bharbar$ is actually more strongly 
related to host-galaxy total stellar mass ($\mstar$) than 
SFR (also see \hbox{\citealt{fornasini18}} for a similar conclusion). 
Their results suggest that the apparent $\bharbar$-SFR 
relation is only a secondary effect resulting from a primary 
$\bharbar$-$\mstar$ relation and the star-formation main sequence.
\hbox{\citet{yang18b}} further show that once $\mstar$ is 
carefully controlled, SMBH accretion is largely independent 
of the cosmic environment of the host galaxies, consistent 
with previous AGN clustering studies 
(e.g. \hbox{\citealt{georgakakis14}}; 
\hbox{\citealt{leauthaud15}}; \hbox{\citealt{mendez16}}; 
\hbox{\citealt{powell18}}).
Motivated by the important role of $\mstar$ in connecting SMBHs
and host galaxies, \hbox{\citet{yang18}} quantitatively 
derived the $\bharbar$-$\mstar$ relation at different redshifts
up to $z=4$.
Aided by the stellar-mass history from \hbox{\citet{behroozi13}},
\hbox{\citet{yang18}} predicted the typical $\mbh$-$\mstar$ 
relation in the local universe. 
At the massive end ($\mstar \gtrsim 10^{11.2}$~$M_\odot$) of 
their $\mbh$-$\mstar$ relation, their $\mbh/\mstar$ ratio is 
$\approx 1/500$, similar to observed $\mbh/\mbulge$ values 
(e.g. \hbox{\citealt{haring04}}; \hbox{\citealt{kormendy13}}).
This agreement is expected, as the bulge becomes dominant 
and $\mbulge \approx \mstar$ for massive galaxies. 

Despite the $\bharbar$-$\mstar$ relation being generally  
supported by observations, it cannot straightforwardly explain 
the tightness of the $\mbh$-$\mbulge$ correlation. 
The key to the origin of the tight $\mbh$-$\mbulge$ correlation might 
be related to the morphology of host galaxies, since $\mbh$
is only correlated with the masses of classical bulges rather 
than other galactic components such as pseudo-bulges or disks
(e.g. \hbox{\citealt{kormendy13}} and references therein).
Therefore, SMBH growth might be related to star formation
activity of the bulge only. 
To investigate this potential SMBH-bulge coevolution, one should 
ideally study the relation between $\bharbar$ and bulge SFR in 
the distant universe.  
However, with current facilities, it is infeasible to separate 
the bulge SFR from total SFR when disks are present.
In this work, we focus on a sample of bulge-dominated galaxies 
for which bulge SFR $\approx$ total SFR. 
If SMBHs indeed coevolve with host-galaxy bulges, we expect
a strong correlation between $\bharbar$ and SFR for these 
bulge-dominated galaxies over cosmic history.

Our bulge-dominated sample is selected from the Cosmic Assembly
Near-Infrared Deep Extragalactic Legacy Survey (CANDELS) where
deep \hst\ $H$-band observations are available 
(\hbox{\citealt{grogin11}}; \hbox{\citealt{koekemoer11}}).
\xray\ emission is a robust tracer of SMBH accretion 
(e.g. \hbox{\citealt{brandt15}} and references therein).
The CANDELS fields also have deep \chandra\ \xray\ observations,
allowing us to estimate reliable $\bharbar$ for any given 
sample of galaxies (e.g. \hbox{\citealt{yang17, yang18b}}).

This paper is structured as follows. 
In \S\ref{sec:analyses}, we describe the data used in this 
work and define our samples. 
In \S\ref{sec:res}, we perform data analyses and present the 
results. 
We discuss our results in \S\ref{sec:discuss}.
We summarize our work and discuss future prospects in 
\S\ref{sec:sum}.

Throughout this paper, we assume a cosmology with 
$H_0=70$~km~s$^{-1}$~Mpc$^{-1}$, $\Omega_M=0.3$, 
and $\Omega_{\Lambda}=0.7$.
We adopt a Chabrier initial mass function (IMF; 
\hbox{\citealt{chabrier03}}).
Quoted uncertainties are at the $1\sigma$\ (68\%)
confidence level. 
We express $\mbulge$, $\mstar$, and $\mbh$
in units of $M_\odot$, SFR and $\bharbar$ in units of 
$M_\odot$~yr$^{-1}$.
$\lx$\ indicates AGN \xray\ luminosity 
at rest-frame \hbox{2--10 keV} and is in units of 
erg~s$^{-1}$. 

\section{Data and Sample}
\label{sec:analyses}
Our analyses are based on the five CANDELS fields, i.e., 
\hbox{GOODS-S}, \hbox{GOODS-N}, EGS, UDS, and COSMOS
\citep{grogin11, koekemoer11}.
All these fields have multiwavelength observations from \hst, 
\spitzer, and ground-based telescopes such as Subaru and VLT.
These high-quality data sets allow for measurements 
of galaxy morphology (\S\ref{sec:morph}), 
stellar mass ($\mstar$; \S\ref{sec:m_sfr}), and star-formation 
rate (SFR; \S\ref{sec:m_sfr}). 
FIR observations from \herschel\ are also available in 
these fields, enabling robust SFR estimation based on cold-dust 
emission (\S\ref{sec:m_sfr}). 
All of the five CANDELS fields have \chandra\ \xray\ observations
from which we derive $\bharbar$ (\S\ref{sec:bhar}).
We define our sample in \S\ref{sec:samp} and the sample properties 
are summarized in Tab.~\ref{tab:sample}.

\begin{table*}
\begin{center}
\caption{Summary of sample properties}
\label{tab:sample}
\begin{tabular}{ccccccccc}\hline\hline
Field & Area & Gal.\ \# & Spec./Photo. \#  & Galaxy Ref. &
B.-D.\ (X) & Comp.\ (X) & X. Dep. & \xray\ Ref. \\ 
(1) & (2) & (3) & (4) & (5) & (6) & (7) & (8) & (9) \\  
\hline
GOODS-S & 170 & 1,504 & 727/777 & \citet{santini15}   & 398 (100) & 1,106 (241) &  7~Ms & \citet{luo17} \\
GOODS-N & 170 & 1,855 & 391/1,464 & Barro et al., in prep.   & 483 (71) & 1,372 (168) &  2~Ms & \citet{xue16} \\
EGS & 200 & 2,446 & 219/2,227 & \citet{stefanon17}   & 591 (48) & 1,855 (105) &  800~ks & \citet{nandra15} \\
UDS & 200 & 2,128 & 254/1,874 & \citet{santini15}   & 549 (42) & 1,579 (75) &  600~ks & \citet{kocevski18} \\
COSMOS & 220 & 2,369 & 10$^*$/2,359 & \citet{nayyeri17}   & 603 (22) & 1,766 (39) &  160~ks & \citet{civano16} \\
\hline
Total & 960 & 10,302 & 1,601/8,701 & --   & 2,624 (283) & 7,678 (628) &  -- & -- \\
\hline 
\end{tabular}
\end{center}
\begin{flushleft}
{\sc Note.} ---
(1) CANDELS field name. 
(2) Field area in arcmin$^2$. 
(3) Number of galaxies in our $\mstar$-complete sample (\S\ref{sec:samp}).
(4) Number of \hbox{spec-$z$}/\hbox{photo-$z$} sources (\S\ref{sec:m_sfr}). 
(5) Reference for CANDELS galaxy catalog. 
(6) \& (7) Sample size of bulge-dominated and comparison galaxies (\S\ref{sec:samp}). 
The number in parentheses means the sample size of \xray\ detected sources.
(8) \xray\ depth in terms of exposure time (\S\ref{sec:bhar}).  
(9) Reference for \xray\ catalog. \\
$*$ Although there are more than 500 spec-$z$ available {in the CANDELS region of COSMOS}, the 
latest version of the {CANDELS/COSMOS} catalog is mostly based on photo-$z$. 
Future releases of the {CANDELS/COSMOS} catalog will adopt spec-$z$ when available 
(H.\ Nayyeri 2018, private communication). 
\end{flushleft}
\end{table*}

\subsection{Morphology}
\label{sec:morph}
Rest-frame optical/NIR light is essential for morphological 
measurements \citep[e.g.][]{conselice14}.
The \hst\ $H$ band, centered at $\approx 1.6$~$\mu$m, can cover 
rest-frame optical/NIR wavelengths up to $z\approx 3$. 
We adopt the $H$-band morphological measurements in 
\citet{huertas_company15} that are based on machine learning 
for CANDELS galaxies with $H<24.5$. 
The machine-learning technique is chosen to approximate 
visual morphologies from humans, and is trained with a 
galaxy sample that has morphological measurements performed 
by human classifiers \citep{kartaltepe15}.
This training sample has the same magnitude cut ($H<24.5$) 
as reliable visual morphological measurements are difficult 
at fainter magnitudes. 
For each galaxy, the \citet{huertas_company15} catalog provides 
five fractional numbers,
i.e., $f_{\rm sph}$, $f_{\rm disk}$, $f_{\rm irr}$, $f_{\rm pt}$,
and $f_{\rm unc}$.
These fractions represent the probabilities that a hypothetical classifier 
would have voted for a galaxy having a spheroid, a disk, and some 
irregularities, being point-like and unclassifiable, respectively. 
Note that the sum of the fractions might exceed unity, because,
for example, a galaxy might have both spheroidal and disky features
simultaneously.

A high $f_{\rm unc}$ value indicates that the source might be 
a spurious detection, e.g. the spikes of a bright star 
being falsely detected as a source (see, e.g. Fig.~13 of 
\hbox{\citealt{huertas_company15}}).
Sources with high $f_{\rm pt}$ value might be stars or broad-line 
(BL) AGNs. 
Due to strong light from the AGN central engine, morphology 
measurements of host galaxies are unreliable for luminous BL AGNs 
(e.g. \S5.3 of \hbox{\citealt{brandt15}}).
We exclude the $\approx 8\%$ of sources that have $f_{\rm unc}$ or 
$f_{\rm pt}$ greater than any of $f_{\rm sph}$, $f_{\rm disk}$, 
and $f_{\rm irr}$. 
Upon visual inspection, the excluded sources are indeed spurious 
detections or point-like. 
Morphological measurements are challenging at high redshift
and our work probes up to $z=3$. 
We discuss some possible redshift-related effects on our results
in \S\ref{sec:reli}.

\subsection{Redshift, Stellar Mass, and Star Formation Rate}
\label{sec:m_sfr}
We obtain redshift measurements from the CANDELS catalogs
(see Tab.~\ref{tab:sample}).
These measurements are spectroscopic redshifts (spec-$z$) or 
photometric redshifts (photo-$z$).
The photo-$z$ measurements are based on dedicated photometry 
extracted with careful consideration of PSF sizes and source 
shapes (e.g. \hbox{\citealt{guo13}}; \hbox{\citealt{galametz13}}).
Compared to the available spec-$z$, the photo-$z$ shows high
quality, with $\sigma_{\rm NMAD} = 0.018$ and an outlier fraction
of 2\%.\footnote{Here, $\sigma_{\rm NMAD}$ is defined as 
$1.48\times \mathrm{median}(\frac{|\Delta z- \mathrm{median}(\Delta z)|}
{1+z_\mathrm{spec}})$, and outliers are defined as sources having 
${|\Delta z|/(1+z_{\rm spec})>0.15}$.}
As in \S\ref{sec:morph}, we discard the 79 spectroscopic BL 
AGNs reported in the literature
(\hbox{\citealt{barger03}}; \hbox{\citealt{silverman10}}; 
\hbox{\citealt{cooper12}}; \hbox{\citealt{newman13}}; 
\hbox{\citealt{marchesi16}}; Suh et al., in prep.).

\begin{figure*}
\includegraphics[width=0.49\linewidth]{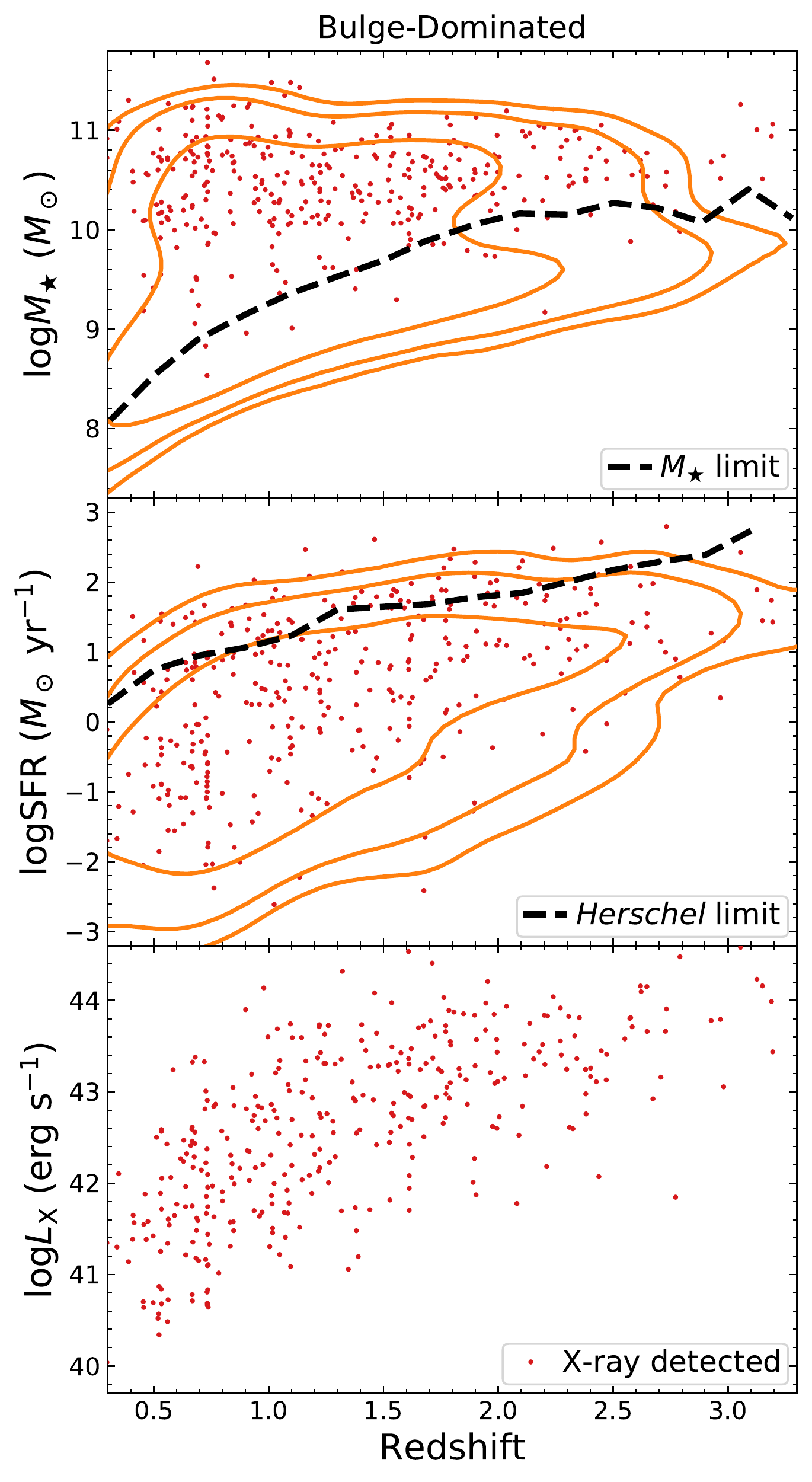}
\includegraphics[width=0.49\linewidth]{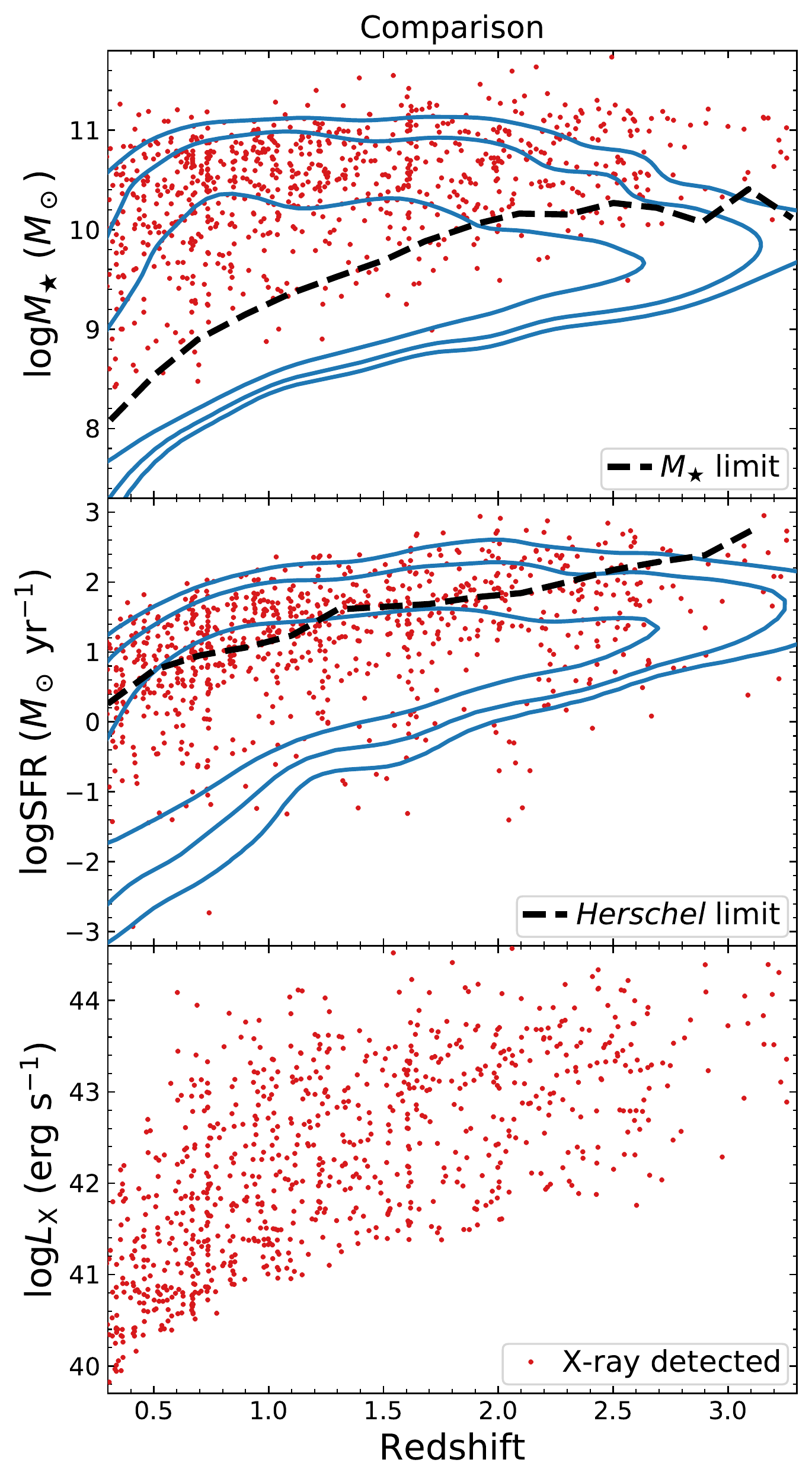}
\caption{$\mstar$, SFR, and $\lx$ as a 
function of redshift for the bulge-dominated (left) and 
comparison (right) samples. 
The contours encircle 68\%, 90\%, and 95\% of all 
$H<24.5$ galaxies, respectively. 
The red points represent \xray\ detected sources.
In the top panels, the dashed curve indicates the 
$\mstar$ completeness limit (\S\ref{sec:m_sfr}).
In the middle panels, the dashed curve indicates the SFR 
values above which 70\% of sources have FIR-based SFR 
(\S\ref{sec:m_sfr}). 
}
\label{fig:all_vs_z}
\end{figure*}

The CANDELS catalogs also provide 
stellar mass ($\mstar$) and star formation rate (SFR) 
measurements from independent teams. 
Following \citet{yang17}, we adopt the median $\mstar$
and SFR values from the five available teams (2a$_\tau$, 
6a$_\tau$, 11$a_\tau$, 13a$_\tau$, and 14a).\footnote{For
\hbox{GOODS-N}, only three teams are available 
(2a$_\tau$, 6a$_\tau$, and 14a) for now.}
Fig.~\ref{fig:all_vs_z} (top) shows $\mstar$ as a 
function of redshift for $H<24.5$ galaxies that have 
morphological measurements (\S\ref{sec:morph}). 
We limit our analyses to a $\mstar$-complete (corresponding 
to $H<24.5$) sample (\S\ref{sec:samp}).
The limiting $\mstar$ ($\mlim$) for $H<24.5$ is also 
displayed in Fig.~\ref{fig:all_vs_z}. 
The $\mlim$-redshift curve is derived based on an empirical 
method (e.g. \hbox{\citealt{ilbert13}}). 
We first divide our sources into narrow redshift bins
with width of $\Delta z=0.2$. 
For each redshift bin, we calculate 
$\log\mlim^{\rm ind}=\log\mstar + 0.4\times (H-24.5)$ for
individual galaxies in the bin.
We then adopt $\mlim$ as the 90th percentile of the 
$\mlim^{\rm ind}$ distribution for the redshift bin.

The CANDELS $\mstar$ and SFR are based on SED fitting of 
rest-frame UV-to-NIR photometry using galaxy templates.
As demonstrated by previous works \citep[e.g.][]{luo10, yang17, kocevski18},
the rest-frame UV-to-NIR light is often predominantly contributed
by galaxy component rather than the AGN component for \xray\ 
AGNs in the CANDELS fields.
Also, we have removed BL AGNs that might have strong AGN components 
in their UV-to-NIR SED. 
Therefore, the AGN SED contribution should not qualitatively affect our 
results (see \S\ref{sec:reli} for other evidence).

The SED-based SFR estimation, which is physically based
on obscuration-corrected UV light, is reliable for 
low-to-moderate levels ($\mathrm{SFR \lesssim 100}\ \MsunYr$) 
of star-formation activity. 
However, it tends to underestimate SFR in the high-SFR
regime, possibly due to strong dust obscuration 
\citep[e.g.][]{wuyts11, whitaker17, yang17}.
To alleviate this issue, we adopt SFR from FIR photometry 
of \herschel\ when available (\hbox{\citealt{lutz11}}; 
\hbox{\citealt{oliver12}}; \hbox{\citealt{magnelli13}}).
{The photometry has been extracted using \spitzer/MIPS 
24~${\mu}$m priors and source-blending issues 
have been carefully addressed.}
The FIR-based SFR is more robust than the SED-based SFR, 
especially in the high-SFR regime \citep[e.g.][]{chen13, yang17}.
Due to limited sensitivity, \herschel\ can only detect
sources with the highest SFR at a given redshift.  
There are five \herschel\ bands available for the CANDELS fields, 
i.e., 100~$\mu$m, 160~$\mu$m, 250~$\mu$m, 350~$\mu$m, and 
500~$\mu$m.
We only utilize robust detections with $\mathrm{S/N>3}$.
We discard the 100~$\mu$m band at redshifts above $z=1.5$, 
because the observed 100~$\mu$m corresponds to rest-frame 
$< 40\ \mu$m which might be 
contaminated by hot-dust emission powered by AGN activity.
We adopt the reddest available \herschel\ band to estimate SFR, since
longer wavelengths are ``freer'' from possible AGN emission. 
We calculate SFR from FIR flux following the procedure 
in \citet{chen13} and \citet{yang17}.
We first derive galaxy total IR luminosity ($L_{\rm IR}$) 
from FIR flux based on the star-forming galaxy templates of 
\citet{kirkpatrick12}.
We adopt the $z\sim 1$ and $z\sim 2$ templates for $z<1.5$
and $z\geq 1.5$ sources, respectively.
We then obtain SFR as 
\begin{equation}\label{eq:sfr}
\begin{split}
\frac{\mathrm{SFR}}{\MsunYr} = 1.09 \times 10^{-10} 
\frac{L_{\rm IR}}{L_\odot}.
\end{split}
\end{equation}
Fig.~\ref{fig:all_vs_z} (middle) shows SFR (based on SED or 
FIR) as a function of redshift for all $H<24.5$ galaxies.

The comparison sample has a higher fraction of FIR-based SFR
measurements than the bulge-dominated sample (32\% vs.\ 9\%),
because the former generally has stronger star-formation
activity than the latter.
To investigate whether this difference in SFR measurements could
bias our results, we have tested cutting our $z=0.5\text{--}1.5$
($z=1.5\text{--}3$) sample at $\mathrm{SFR}<10$~$M_\odot$~yr$^{-1}$
($\mathrm{SFR}<10^{1.5}$~$M_\odot$~yr$^{-1}$), below which the
SFR measurements are mostly SED-based (see Fig.~\ref{fig:all_vs_z}).
Under these cuts, our results do not change qualitatively.\footnote{We cannot
perform a similar test for high-SFR galaxies with \herschel\ detections,
because the sample size of \herschel-detected sources is too small.}
We have also tested our results using SED-based SFR only for the entire
galaxy sample and our conclusions still hold.
{It is well known that SFR measurements from SEDs and the FIR do not 
always agree \citep[e.g.][]{buat10, rodighiero14}, and \cite{yang17} found 
that the statistical scatter between these two methods is
${\lesssim 0.5}$~dex.
This level of scatter is unlikely to be seriously problematic to 
our statistical analyses (\S\ref{sec:sfr_or_m}), since our SFR bin sizes 
are typically ${\gtrsim 0.5}$~dex. 
To verify this point, we perturb our SFR measurements by 0--0.5~dex 
randomly and our results below in \S\ref{sec:res} do not change qualitatively 
after the perturbation. 
\cite{yang17} found the systematic offset between SED-based and FIR-based 
SFRs is typically small for the general galaxy population 
(${\approx 0.2}$~dex), and this level of systematic should not 
change our main conclusions considering our relatively large SFR bin sizes.
However, \cite{symeonidis16} considered that, for strong AGNs, FIR-based 
SFR might be systematically overestimated due to the contamination of 
AGN-heated dust. 
To assess this potential issue, we test our results by using SED-based 
SFR only for strong AGNs (${\log\lx > 43.5}$), and our results 
below do not change qualitatively.
Therefore, our main conclusions should be robust against uncertainties of 
SFR measurements.}

\subsection{The Bulge-Dominated and Comparison Samples}
\label{sec:samp}
Our analyses are based on a bulge-dominated sample and 
a comparison sample. 
In this Section, we detail the selections of these two 
samples. 
We first select all $H<24.5$ galaxies for which morphology 
measurements are available (\S\ref{sec:morph}).
As in \S\ref{sec:morph} and \S\ref{sec:m_sfr}, we 
exclude BL AGNs, stars, and false detections. 
We then divide these galaxies into two redshift bins, i.e., 
$z=0.5\text{--}1.5$ and $z=1.5\text{--}3.0$ for our 
analyses.
The relatively broad redshift bins are necessary to guarantee
sufficiently large samples for our statistical analyses 
(\S\ref{sec:res}). 
We have also tested on narrower redshift bins and found our
qualitative results do not change, although the statistical 
scatter becomes larger due to reduced sample sizes.
Therefore, the two wide redshift bins 
($z=0.5\text{--}1.5$ and $z=1.5\text{--}3.0$) should not 
bias our results, and we adopt them throughout this paper. 

We select $\mstar$-complete samples for the two 
redshift bins.
The limiting $\mstar$ at $z=1.5$ and $z=3.0$ are 
$\log\mstar\approx 9.7$ and $\log\mstar \approx 10.2$, 
respectively (see Fig.~\ref{fig:all_vs_z}). 
Therefore, we limit our analyses to $\log\mstar> 9.7$
and $\log\mstar> 10.2$ galaxies for the low and high 
redshift bins, respectively. 
These $\mstar$ thresholds are below the characteristic $\mstar$ of 
the stellar-mass function (SMF), i.e., $\log \mstar \approx 10.6$
at $z\approx 0.5\text{--}3$ \citep[e.g.][]{tomczak14, davidzon17}. 
The stellar-mass density above these $\mstar$ cuts is 
$\approx 90\%$ and $\approx 70\%$ of the total for the low
and high redshift bins, respectively (calculated with the
SMF in \hbox{\citealt{behroozi13}}). 
After applying the $\mstar$ cuts, our sample does not include 
dwarf galaxies ($\log\mstar \lesssim 9.5$). 
Aside from technical constraints, the exclusion of dwarf galaxies 
is also motivated by our major science goal, i.e. investigating the 
origin of the $\mbh$-$\mbulge$ relation.
Since the $\mbh$-$\mbulge$ relation is mostly established for 
$\log \mbulge \gtrsim 10$ \citep[e.g.][]{kormendy13}, we should also
focus on relatively massive galaxies rather than dwarf galaxies.

The basic properties of the $\mstar$-complete sample are 
summarized in Tab.~\ref{tab:sample}.
In the $\mstar$-complete sample, we classify a source as 
bulge-dominated if it satisfies 
$f_{\rm sph}\geq 2/3$, $f_{\rm disk}<2/3$, and $f_{\rm irr}<1/10$. 
These empirical criteria are suggested by several previous 
studies (e.g. \hbox{\citealt{huertas_company15b, huertas_company16}}; 
\hbox{\citealt{kartaltepe15}}).
{As indicated by these criteria, the term ``bulge-dominated''
refers to galaxies that only display a significant spheroidal 
component, without obvious disky and/or irregular components.
We note that different authors may adopt different terminology for 
the bulge-dominated galaxies (e.g. ``spheroid-like''; e.g. 
\citealt{conselice14}).}
If a galaxy does not meet these criteria, we include it in our 
comparison sample, i.e., galaxies that are not bulge-dominated.
The bulge-dominated and comparison samples have $\approx$~2,600
and 7,700 galaxies, respectively (see Tab.~\ref{tab:sample}).
Our analyses in \S\ref{sec:res} are based on these two samples. 

\begin{figure*}
\includegraphics[width=0.49\linewidth]{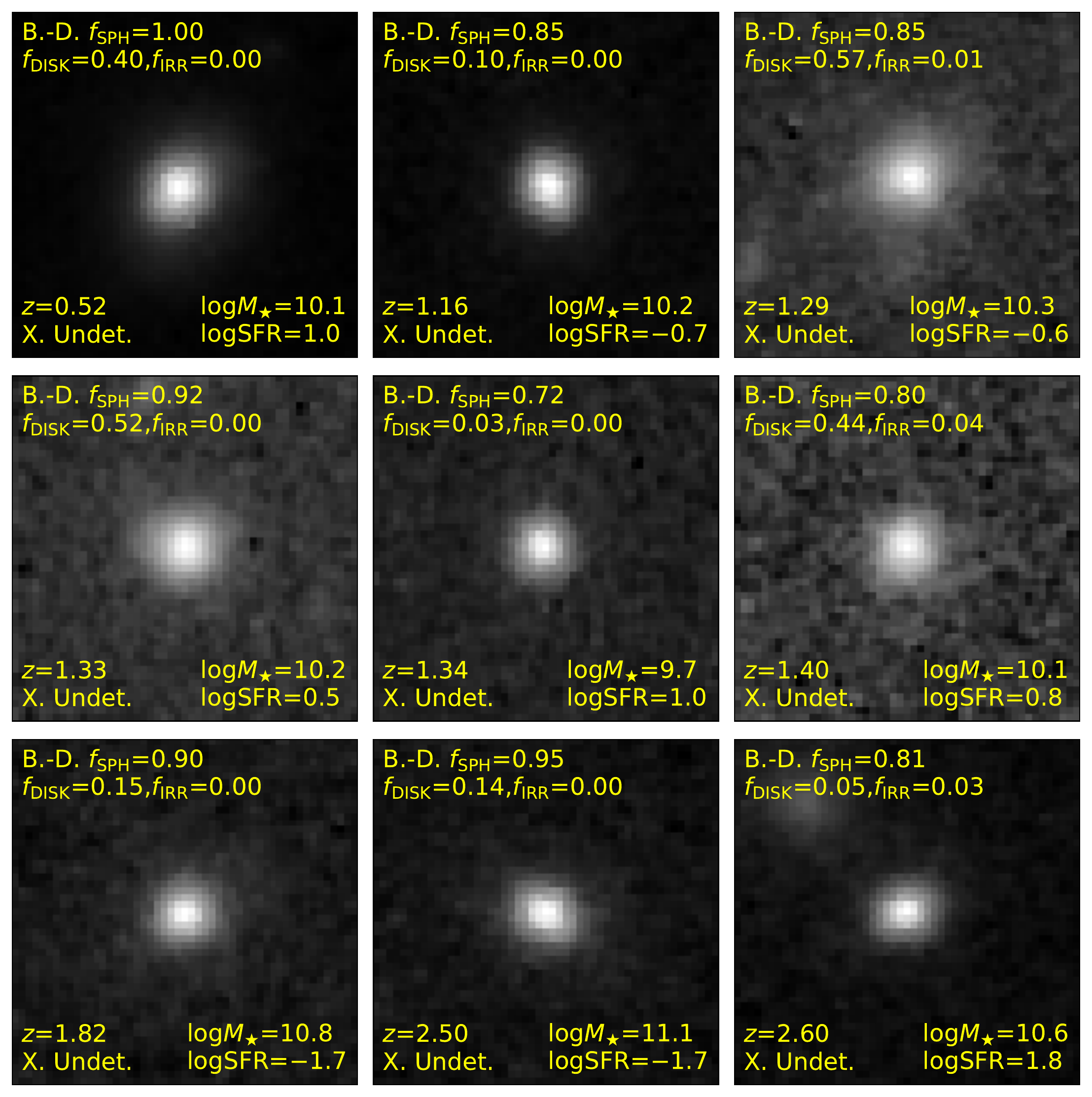}
\includegraphics[width=0.49\linewidth]{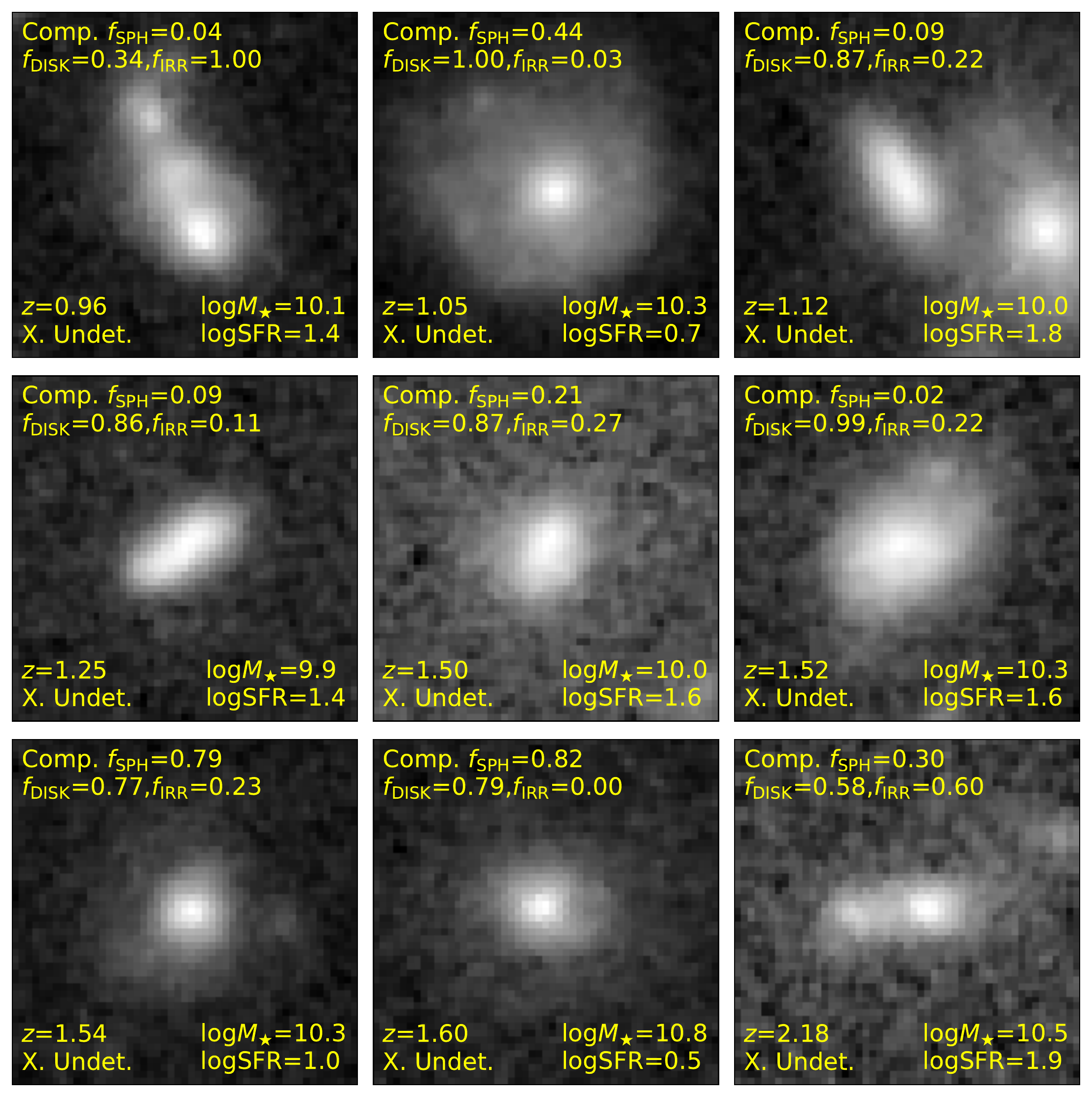}
\includegraphics[width=0.49\linewidth]{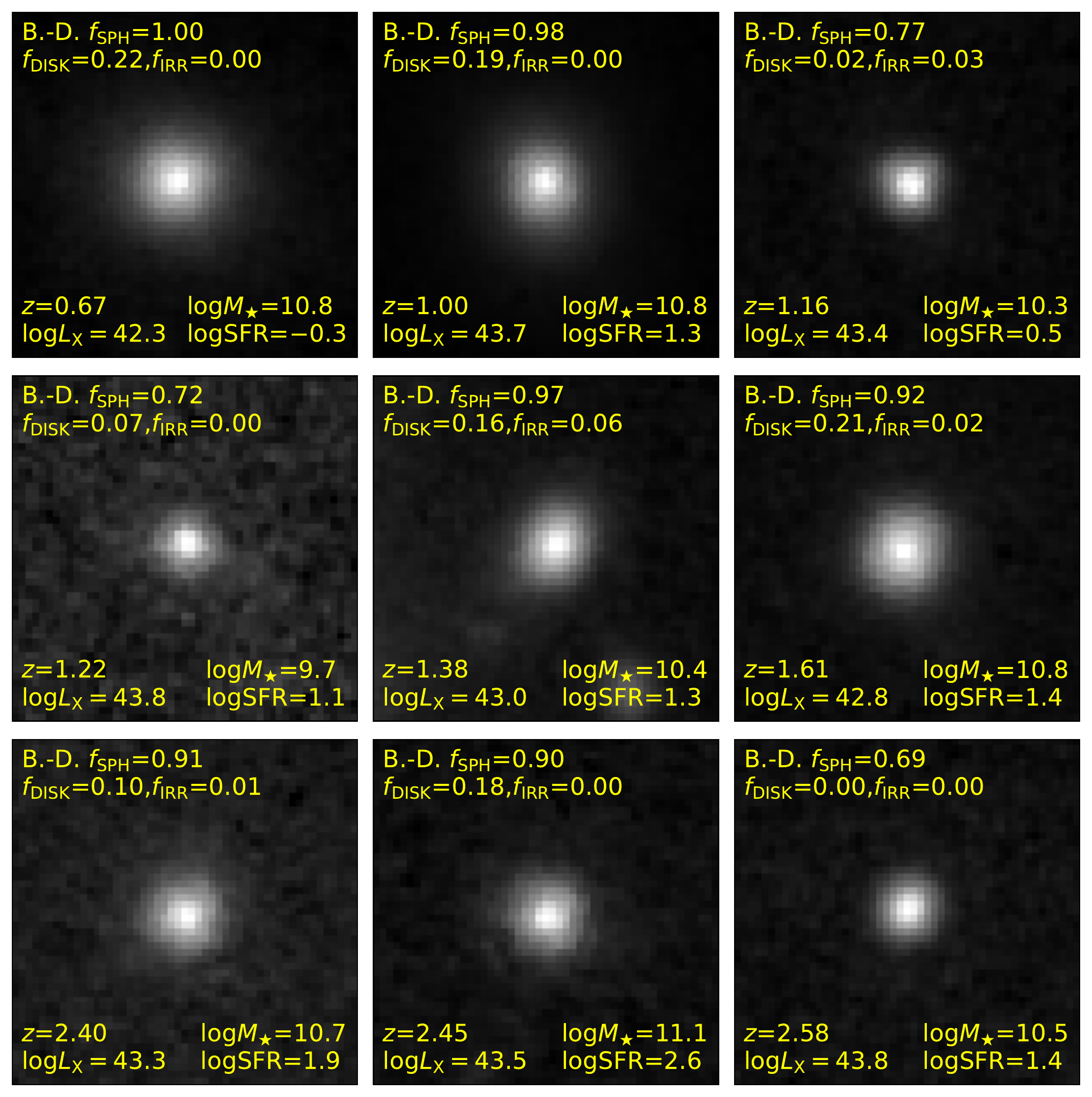}
\includegraphics[width=0.49\linewidth]{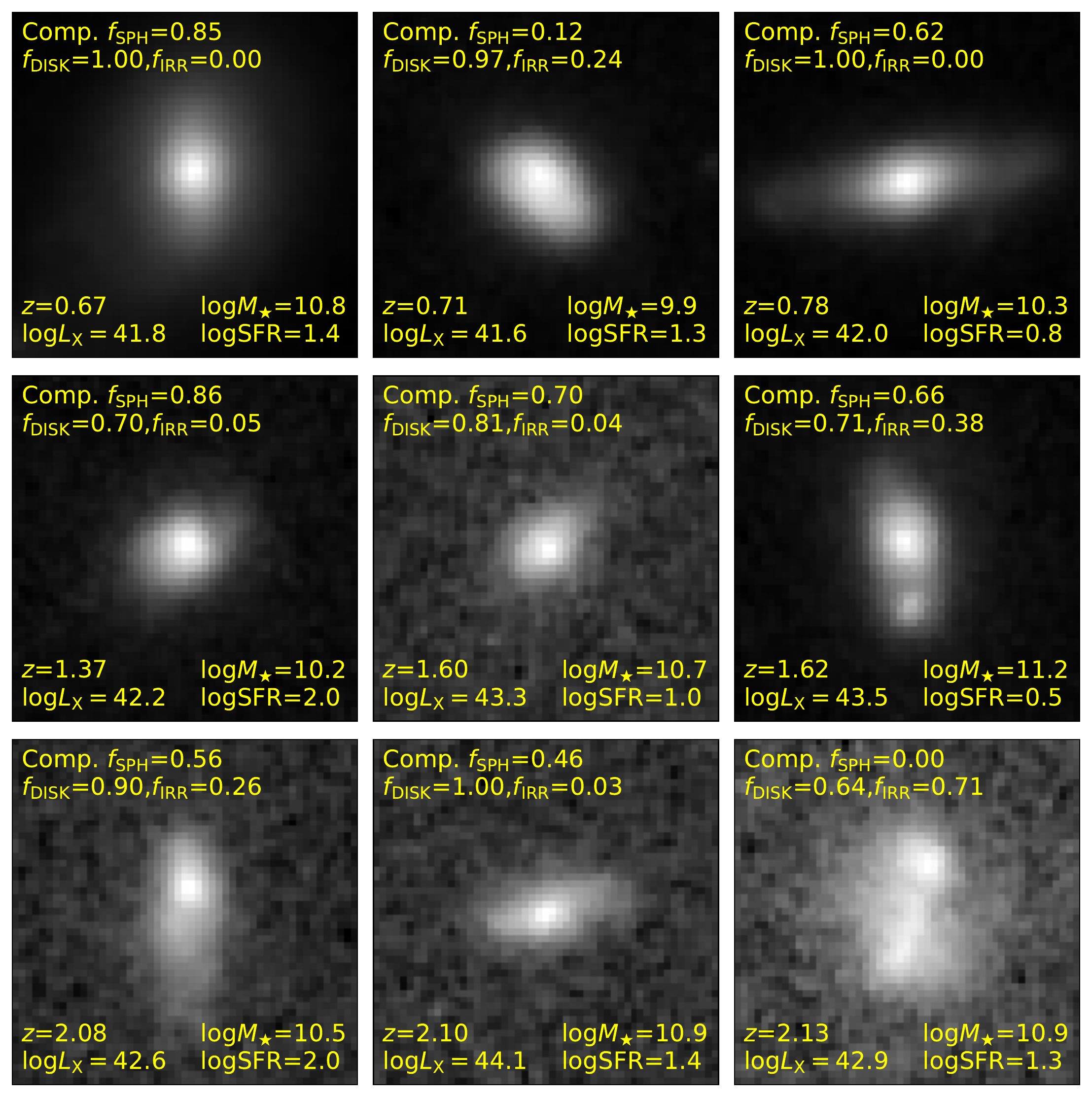}
\caption{Example $H$-band $3''\times3''$ cutouts for
    the bulge-dominated (left) and comparison (right) samples,
    and the \xray\ undetected (top) and detected (bottom) samples.
    {For each sample, the cutouts are arranged in ascending
    order of redshift.}
    The galaxy of interest is placed at the center of each cutout.
    These galaxies are randomly selected from our sample in 
    \S\ref{sec:samp}.
    Note that galaxies can simultaneously have high $f_{\rm sph}$
    and $f_{\rm disk}$ values;
    these galaxies are not selected as bulge-dominated and are 
    included in the comparison sample (see \S\ref{sec:samp}).
}
\label{fig:thumb}
\end{figure*}

\begin{figure*}
\includegraphics[width=0.49\linewidth]{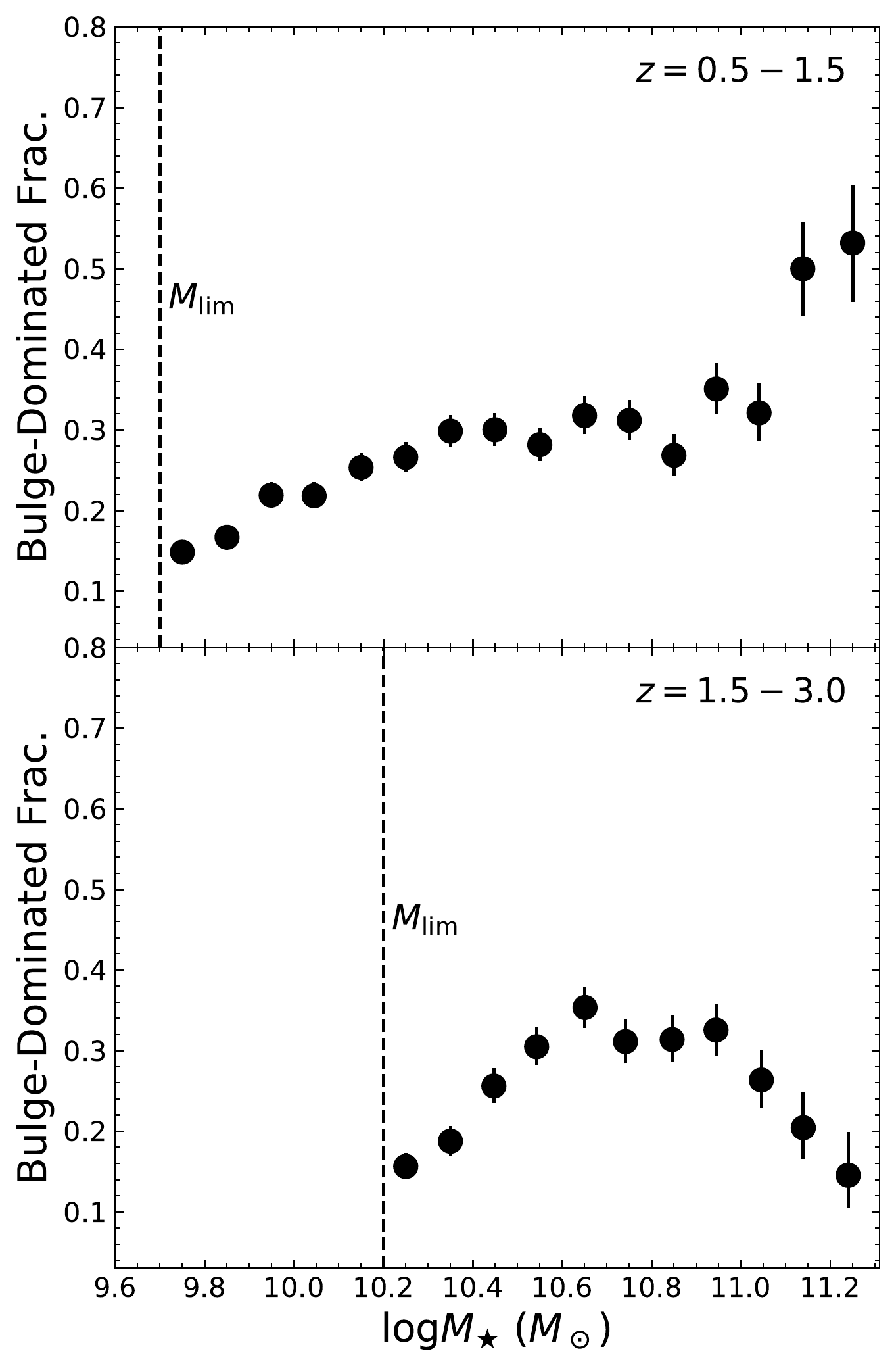}
\includegraphics[width=0.49\linewidth]{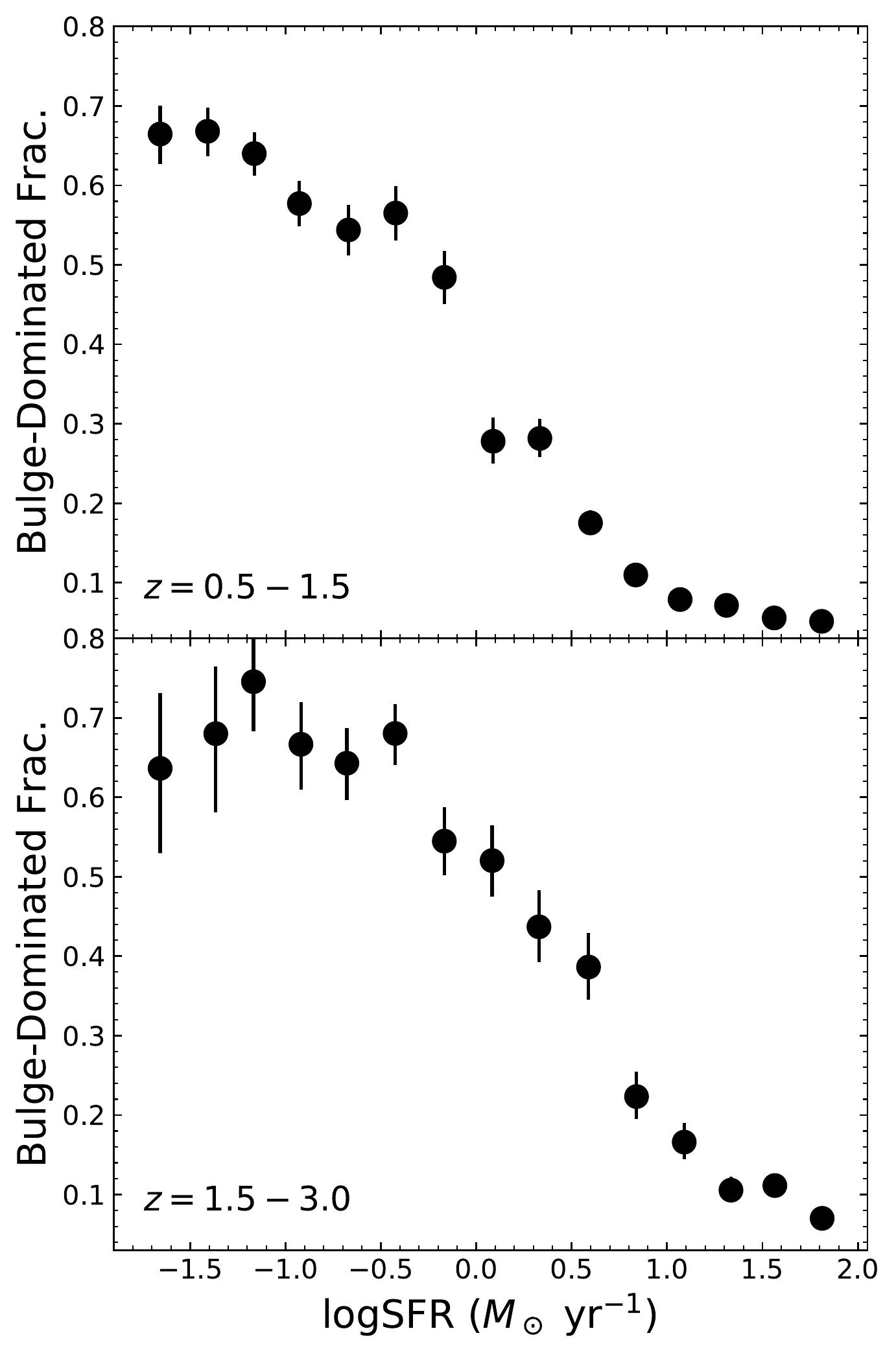}
\caption{The fraction of bulge-dominated galaxies as a function of 
$\mstar$ (left) and SFR (right).
The error bars represent binomial uncertainties.
The vertical dashed lines indicate the limiting $\mstar$ corresponding 
to $H<24.5$ (\S\ref{sec:m_sfr}).
}
\label{fig:sph_frac}
\end{figure*}

In Fig.~\ref{fig:thumb}, we show some random $H$-band cutouts for
the bulge-dominated and comparison samples, respectively.  
The fractions of bulge-dominated galaxies are both $\approx 25\%$ 
for the low and high redshift bins.
Fig.~\ref{fig:sph_frac} shows the fraction of bulge-dominated galaxies 
as a function of $\mstar$ and SFR, respectively.
At the high-$\mstar$ end ($\log\mstar \gtrsim 11$), the 
bulge-dominated fraction in the low-redshift bin is much higher
than that in the high-redshift bin ($\approx 50\%$ vs.\ $\approx 20\%$).
This is probably due to the fact that galaxy mergers/interactions for 
massive galaxies are increasingly prevalent toward high redshift, and 
thus galaxy irregularities are much stronger toward the early universe 
\citep[e.g.][]{conselice14, marsan18}. 
The bulge-dominated fraction drops significantly toward high SFR, 
indicating that bulge-dominated galaxies tend to have low SFR. 
Similar trends have also been found in previous studies (e.g. 
\hbox{\citealt{huertas_company15b, huertas_company16}}). 
The underlying physical reason might be ``morphological quenching'', 
such that bulges can effectively suppress star formation 
\citep[e.g.][]{martig09}.

\begin{figure*}
\includegraphics[width=0.49\linewidth]{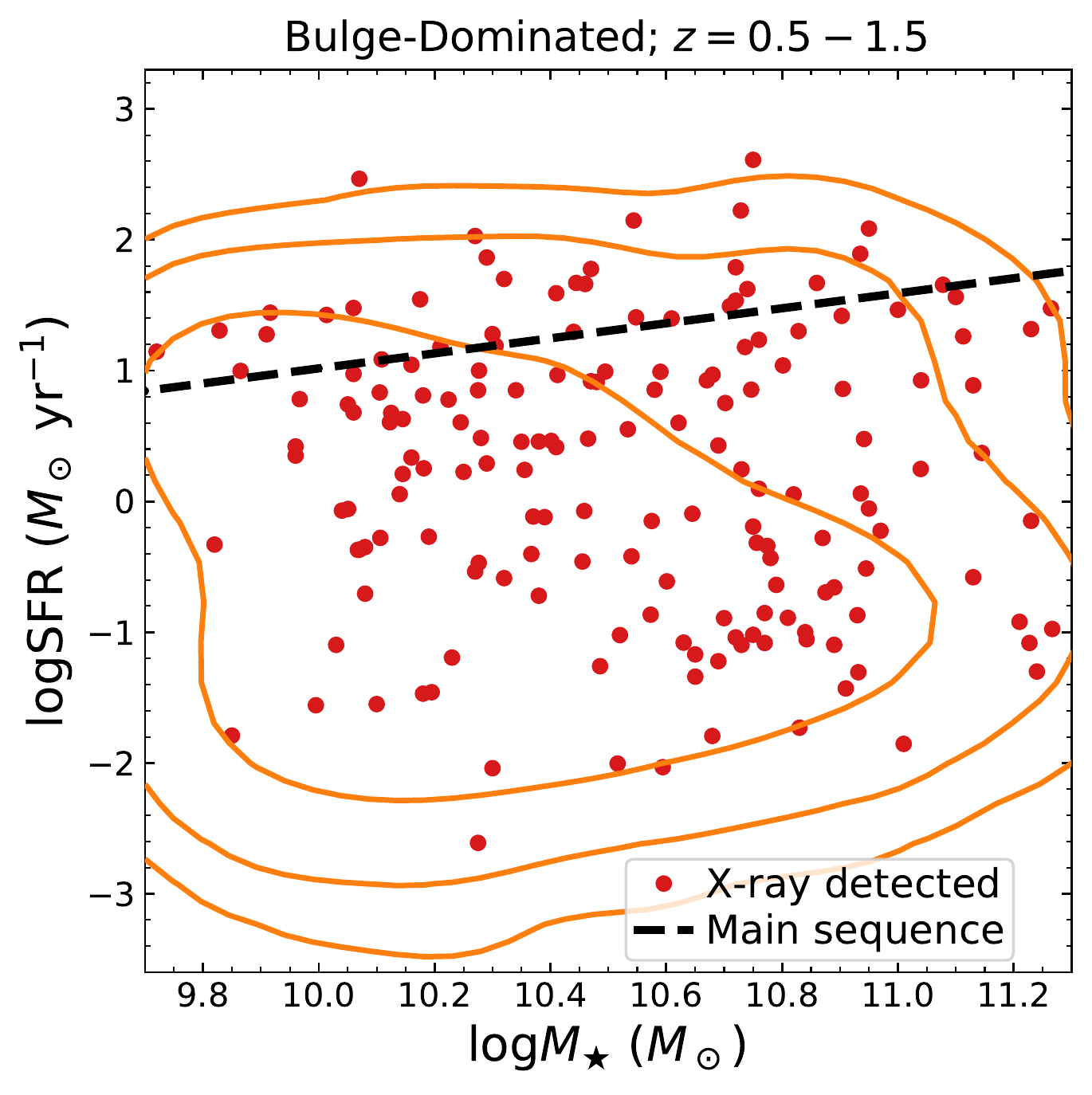}
\includegraphics[width=0.49\linewidth]{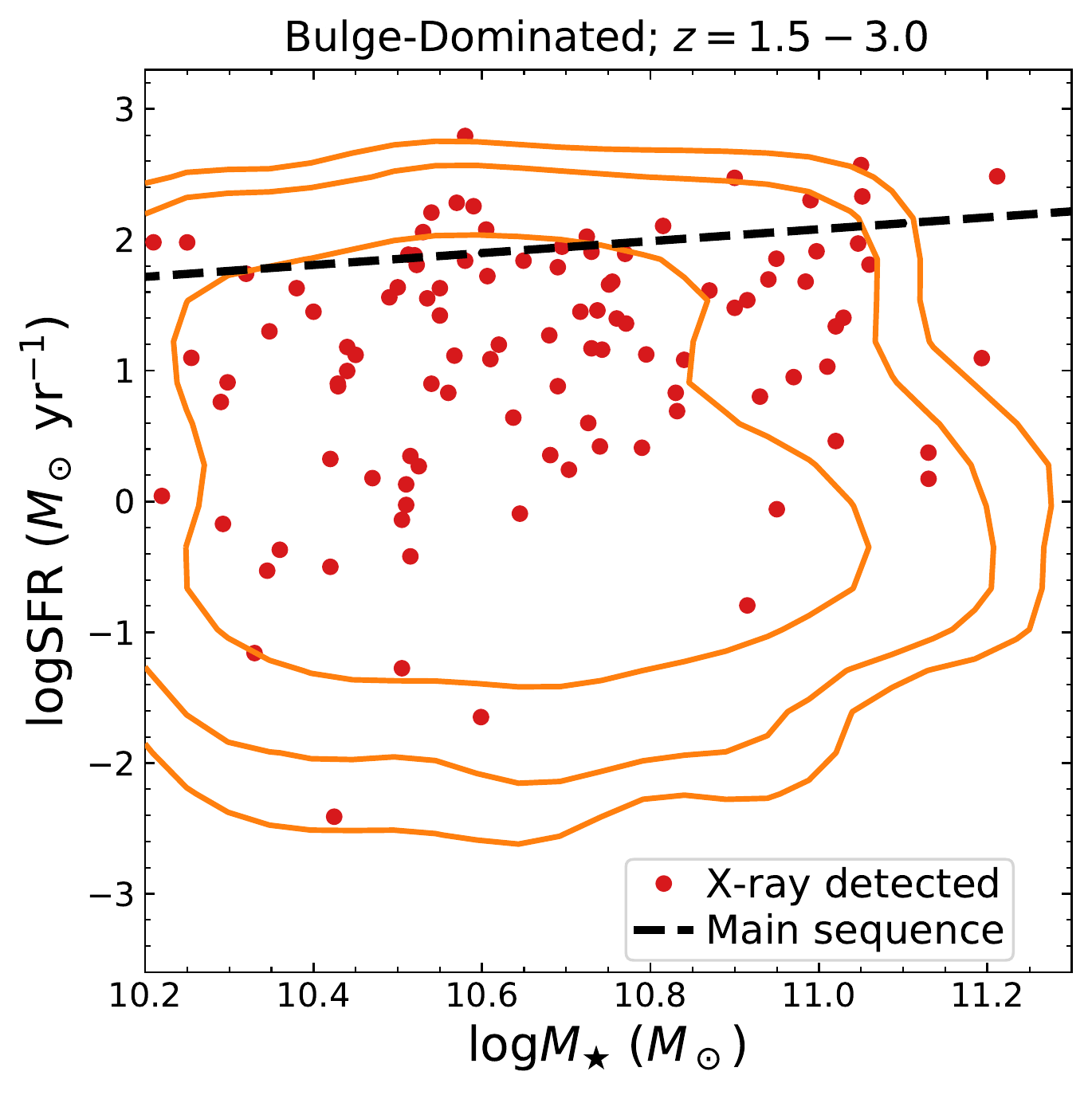}
\includegraphics[width=0.49\linewidth]{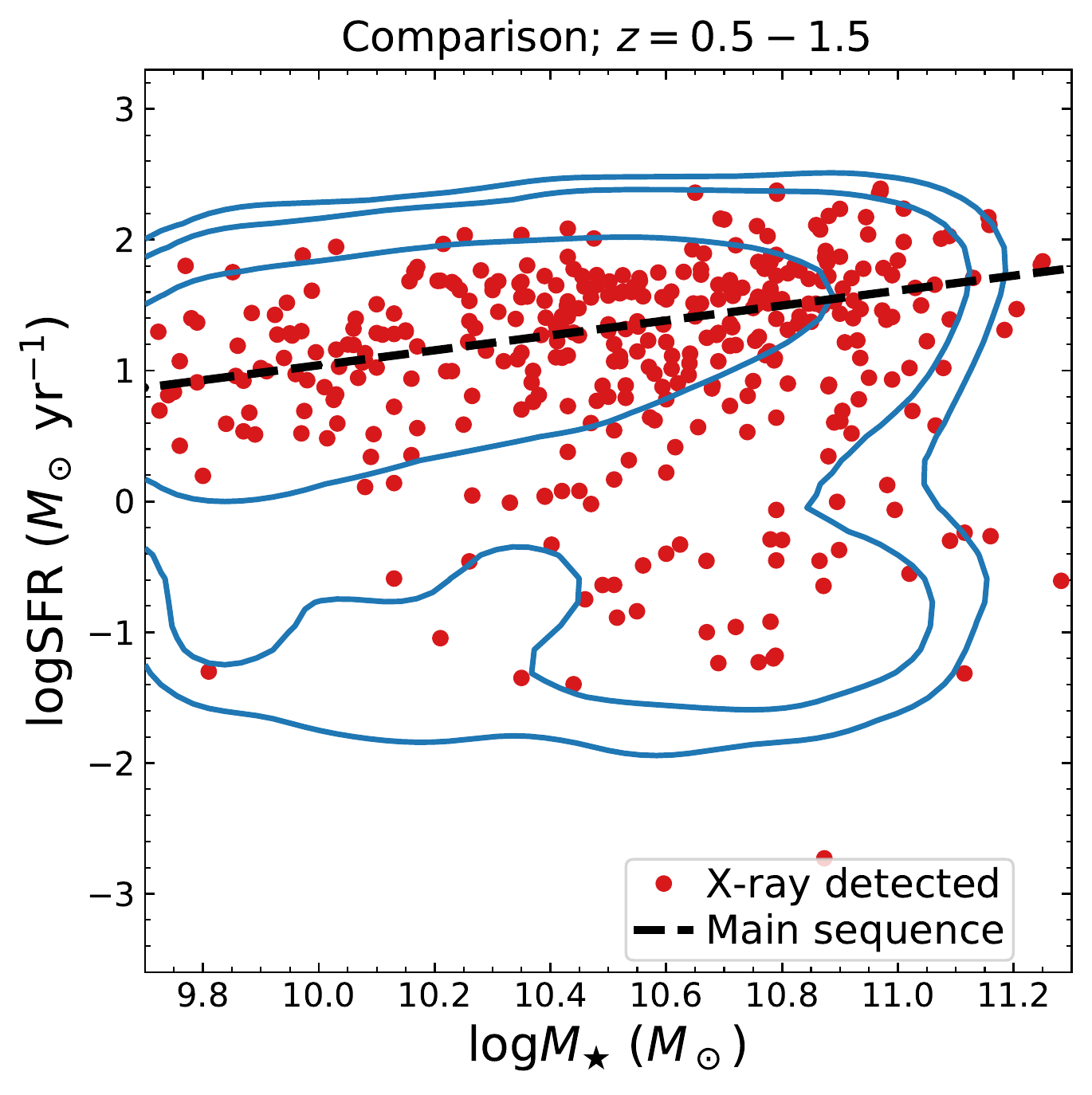}
\includegraphics[width=0.49\linewidth]{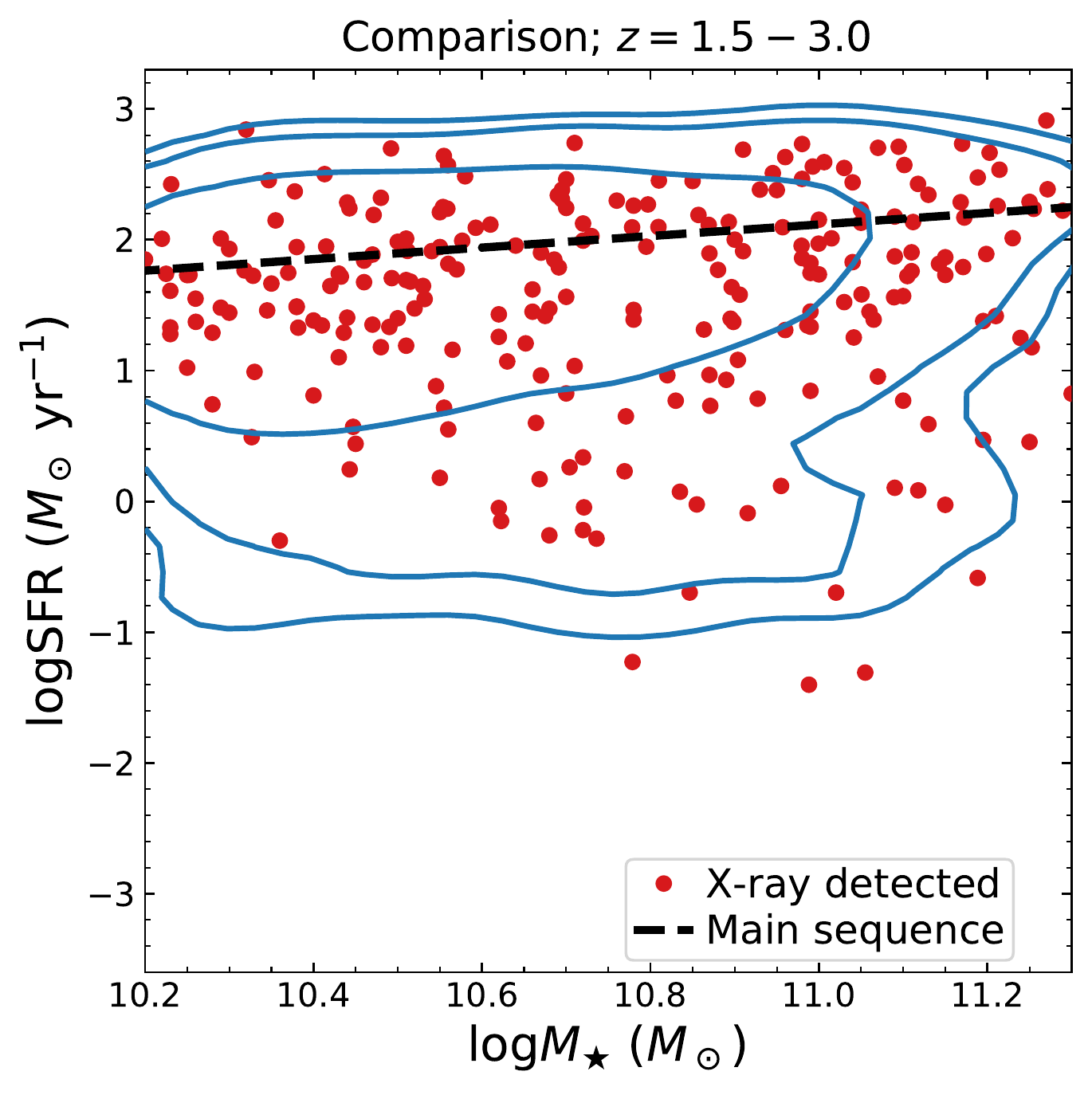}
\caption{The SFR-$\mstar$ distribution for the bugle-dominated (top) 
and comparison (bottom) samples for $z=0.5\text{--}1.5$ (left) 
and $z=1.5\text{--}3.0$ (right). 
The contours encircle 68\%, 90\%, and 95\% of sources, respectively.
The red points represent \xray\ detected sources.
The dashed lines indicate the star-formation main sequence 
at $z=0.98$ (left) and $z=1.97$ (right), respectively 
\hbox{\citep{whitaker12}}. 
$z=0.98$ and $z=1.97$ are the median redshifts for our sources at 
$z=0.5\text{--}1.5$ and $z=1.5\text{--}3.0$, respectively. 
The bulge-dominated sample tends to have lower SFR than  
the comparison sample.
}
\label{fig:SFR-M}
\end{figure*}

Fig.~\ref{fig:SFR-M} displays the source distributions on the 
SFR-$\mstar$ plane for the bulge-dominated and comparison samples,
respectively. 
The bulge-dominated sample tends to lie below the star-formation 
main sequence, while the majority of the comparison sample appears 
to be on the main sequence. 
However, we note that our morphological classification is 
essentially different from a star-forming vs.\ quiescent classification.
For example, the quiescent population in our sample is made up of
$\approx 55\%$ bulge-dominated galaxies and $\approx 45\%$
comparison galaxies.
{While the main population of the comparison sample lies on the
main sequence, there is a non-negligible fraction 
(${\approx 20\%}$) of comparison galaxies lying significantly 
(${\approx 1}$~dex) below the main sequence.
We have visually checked the \hst\ cutouts of these low-SFR sources,
and found they appear to have significant disk/irregular components.
Therefore, the existence of such a low-SFR population among our comparison 
sample is likely intrinsic, and does not appear to be caused by 
morphological misclassification.
}

\subsection{Black Hole Accretion Rate}
\label{sec:bhar}
All five CANDELS fields have deep \xray\ observations from
\chandra.
Tab.~\ref{tab:sample} lists the \xray\ depth and number of 
\xray\ detected sources for each CANDELS field.
We calculate $\bharbar$ contributed by both \xray\ detected
and undetected sources, and thus the resulting $\bharbar$
should cover essentially all SMBH accretion.
This procedure allows us to seamlessly analyze all sources 
in different CANDELS fields which have different \xray\ depths.
We have also repeated our analyses but without sources in 
COSMOS, which has \xray\ depth much shallower than other 
fields (see Tab.~\ref{tab:sample}), and our results do not 
change qualitatively.
For each \xray\ detected source, we calculate $\lx$ from
the \xray\ flux from the corresponding \xray\ catalog 
assuming a photon index of $\Gamma=1.7$
(e.g. \hbox{\citealt{yang16}}; \hbox{\citealt{liu17}}).
Following \citet{yang18b}, we choose, in order of priority, 
hard-band (observed-frame $2\text{--}7$~keV),
full-band (observed-frame $0.5\text{--}7$~keV), or 
soft-band (observed-frame $0.5\text{--}2$~keV)
flux to minimize \xray\ obscuration effects. 
Indeed, \hbox{\citet{yang18b}} estimated that, under this 
scheme of band choice, the \xray\ flux decrease due to 
obscuration is typically small ($\approx 20\%$) for bright 
sources in \cdfs, for which there are enough photons to 
assess obscuration.
We increase the \xray\ fluxes of our \xray\ sources by 20\% to
account for the average systematic effect from obscuration.

For \xray\ undetected sources, we employ a stacking technique 
to include their \xray\ emission. 
We perform this process on full-band \xray\ images.\footnote{For 
the EGS field, we use the \xray\ image from \citet{goulding12}, 
since \citet{nandra15} did not produce the \xray\ image for the
entire EGS field.}  
We generally follow the steps in \citet{vito16}, and we 
briefly summarize this procedure below. 
First, we mask \xray\ detected sources in the \xray\ images.
We choose masking radii ($R_\mathrm{msk}$) of $2\times R_\mathrm{90}$, 
$2.25\times R_\mathrm{90}$, and $2.5\times R_\mathrm{90}$
for sources with net counts $<100$, $100 \text{--}1000$, 
and $>1000$, respectively. 
Here, $R_{90}$ means the radius for a 90\% encircled-energy
fraction (EEF), and it is a function of off-axis angle 
(see Appendix~A of \hbox{\citealt{vito16}}).
After source masking, we derive the net count rate for \xray\
undetected sources. 
To enhance signal-to-noise, we adopt an aperture radius 
$R_\mathrm{aper}=R_\mathrm{80}$, $R_\mathrm{75}$, $R_\mathrm{60}$, 
and $R_\mathrm{40}$ for sources with off-axis angle $<3.5\arcmin$, 
$3.5\arcmin\text{--}4.25\arcmin$, $4.25\arcmin\text{--}5.5\arcmin$, 
and $5.5\arcmin\text{--}7.8\arcmin$, respectively. 
We discard sources whose off-axis angle is $>7.8\arcmin$ and/or 
whose apertures overlap with masked regions. 
The counts in the apertures include contributions from both sources
and background, and we need to subtract the background counts. 
We estimate the background counts in an annulus with inner and 
outer radii of $1.1R_{90}$ and $1.1R_{90}+10\arcsec$, respectively.
Due to the limited aperture size, the net counts encircled in the 
aperture only represent a fraction of the total net counts. 
We perform an aperture correction for each source, depending on
the aperture size adopted. 
For example, if $R_\mathrm{aper}=R_\mathrm{80}$ for a source, 
we divide the aperture net counts by 80\% to recover the total 
net counts. 
We then obtain the count rate by dividing the total net 
counts by the exposure time at the position of the source. 
For sources in each field, we derive fluxes by multiplying 
the count rates by a constant factor, which is the median 
flux/count-rate ratio of \xray\ detected sources in the field.
Finally, for a group of \xray\ undetected sources, we can 
obtain their average \xray\ luminosity 
($\overline{L_\mathrm{X, stack}}$) from the average 
\xray\ flux and redshift, assuming $\Gamma=1.7$.
Our derived SMBH accretion power is mostly contributed by 
the \xray\ detected sources, and the stacking procedure 
typically accounts for less than 20\% of the accretion 
power.

For \xray\ detected objects, we have $\lx$ for individual 
sources; for \xray\ undetected sources, we have 
$\overline{L_\mathrm{X, stack}}$ for any group of sources.
We can then calculate average AGN 
bolometric luminosity for any sample of sources as
\begin{equation}\label{eq:lxbar}
\begin{split}
\overline{\lbol} = \frac{ \Sigma_{\rm det} (\lx-L_{\rm X,XRB})\kbol  + 
   (\overline{L_{\rm X,stack}} - \overline{L_{\rm X,XRB}}) 
	N_{\rm non} \overline{\kbol}
}{N_{\rm det} + N_{\rm non}}
\end{split}
\end{equation} 
Here, $N_{\rm det}$ and $N_{\rm non}$ are the numbers
of \xray\ detected and undetected sources in the sample.
$L_{\rm X,XRB}$ is the expected luminosity from \xray\ 
binaries (XRBs) and $\overline{L_{\rm X,XRB}}$ is the 
average XRB luminosity for the stacked sources. 
To obtain $L_{\rm X,XRB}$ and $\overline{L_{\rm X,XRB}}$, 
we adopt model 269 of \citet{fragos13} which describes 
XRB \xray\ luminosity as a linear function of $\mstar$ and SFR. 
Model 269 is a theoretical model favored by the
observations of galaxies at $z=0\text{--}2$
\citep{lehmer16}.
The expected \xray\ emission from XRBs only accounts for 
$\approx 15\%$ of the total \xray\ power, and thus the uncertainties
related to the XRB modelling should not affect our analyses 
significantly.
$\kbol$ and $\overline{\kbol}$ are the $\lx$-dependent bolometric 
corrections at $(\lx-L_{\rm X,XRB})$ and 
$(\overline{L_{\rm X,stack}} - \overline{L_{\rm X,XRB}})$, 
respectively. 
We adopt the bolometric-correction model from 
\citet{hopkins07}.\footnote{As pointed out in Footnote 4 of
\citet{merloni13}, the $\kbol$ in \citet{hopkins07} appears
to be overestimated due to the double counting of IR 
reprocessed emission. Following \citet{merloni13}, we multiply
the $\kbol$ in \citet{hopkins07} by a factor of 0.7 to 
address this issue.}
Assuming a constant radiative efficiency of $\epsilon=0.1$, we 
can convert $\overline{\lbol}$ to $\bharbar$ as 
\begin{equation}\label{eq:bhar}
\begin{split}
\bharbar 
    & = \frac{ (1-\epsilon) \overline{\lbol} }{\epsilon c^2} \\
    & = \frac{1.58 \overline{\lbol}}{10^{46}\ \rm{erg~s^{-1}}}
	    M_{\sun}\ \mathrm{yr}^{-1},
\end{split}
\end{equation}
where $c$ is the speed of light.
The adopted $\epsilon=0.1$ is motivated by observations 
(see, e.g.\ \S3.4 of \citealt{brandt15}).
We obtain the $\bharbar$ uncertainties with a bootstrapping 
technique (e.g. \S2.3 of \citealt{yang17}).

As explained in \S\ref{sec:intro}, the $\bharbar$ quantity is 
designed to approximate long-term average SMBH accretion rate,
and has been widely adopted in the studies of AGN-galaxy 
relations (e.g. \citealt{chen13}; \citealt{hickox14}; 
\citealt{yang17, yang18, yang18b}; \citealt{fornasini18}).
Some works proposed to recover the full distribution of 
BHAR as a function of galaxy properties (e.g. \citealt{volonteri15}; 
\citealt{georgakakis17}; \citealt{aird18, aird18b}), and 
quantities such as $\bharbar$ and duty cycle can then be derived.
However, detailed modelling of the BHAR distribution at given 
$\mstar$, SFR, and morphological type is beyond 
the scope of this work, and we leave it to future studies.

\section{Analyses and Results}
\label{sec:res}
In this Section, we study $\bharbar$ as a function of SFR 
and $\mstar$ (\S\ref{sec:bhar_vs_sfr_m}).
We address the question of whether $\bharbar$ is mainly related to 
SFR or $\mstar$ in \S\ref{sec:sfr_or_m}. 
All these analyses are performed for the bulge-dominated and 
comparison samples, respectively.
In Appendix~\ref{app:all}, we perform the same analyses for 
all galaxies. 
In \S\ref{sec:quan}, we quantify the $\bharbar$-SFR relation 
for the bulge-dominated sample.

\subsection{BHAR as a Function of SFR and $\mstar$}
\label{sec:bhar_vs_sfr_m}
\begin{figure*}
\includegraphics[width=0.49\linewidth]{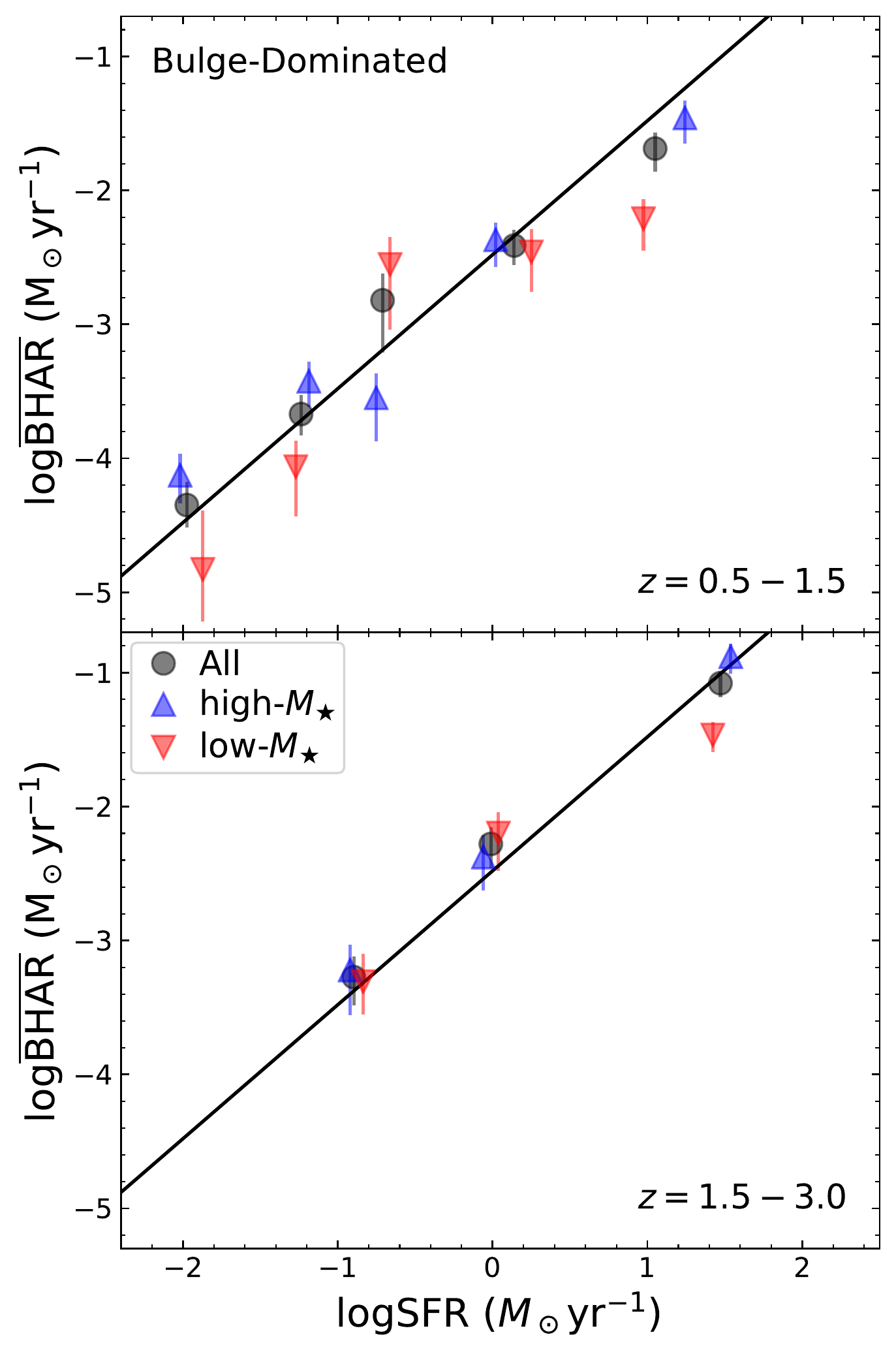}
\includegraphics[width=0.49\linewidth]{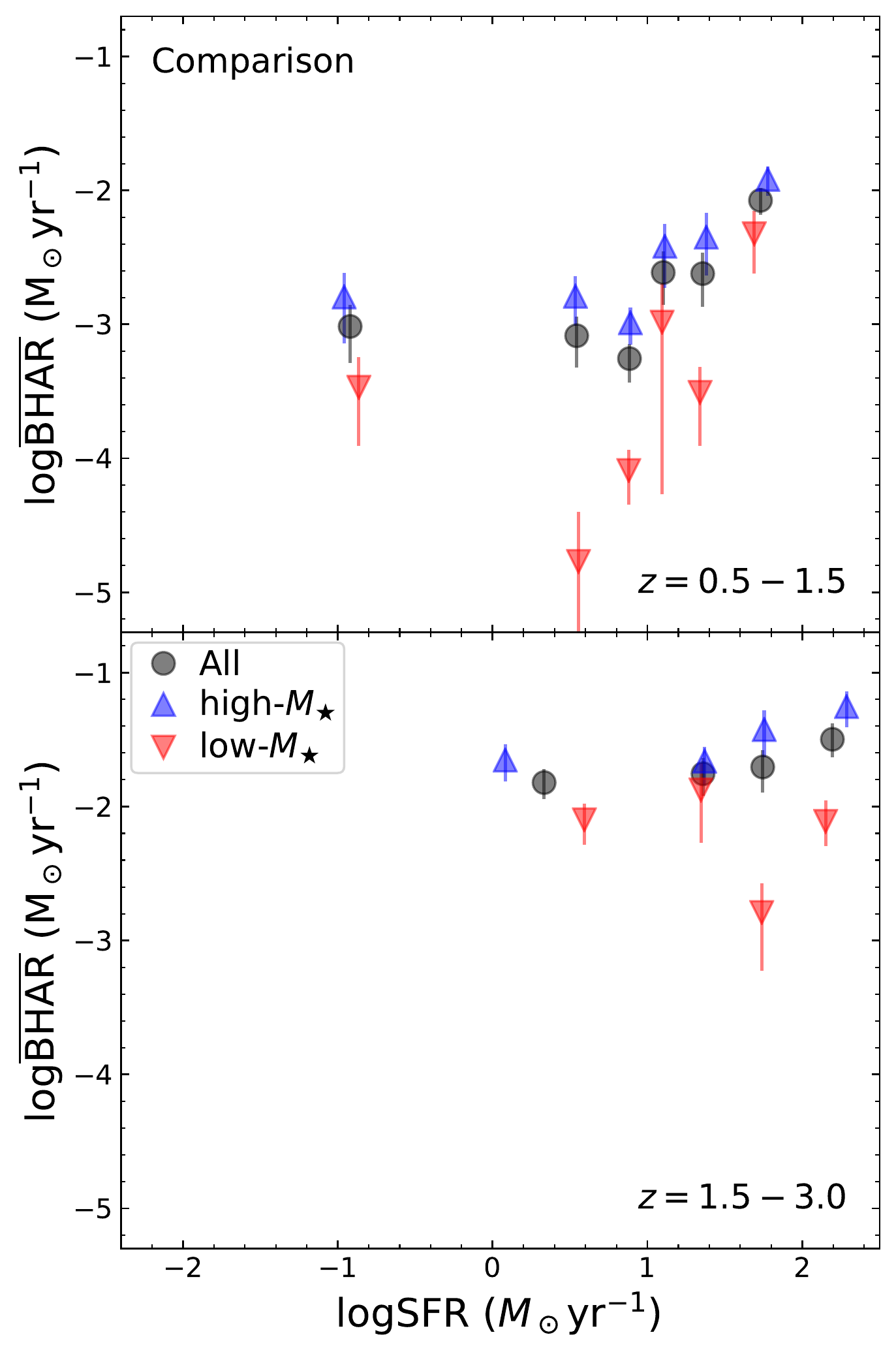}
\caption{$\bharbar$ vs.\ SFR for bulge-dominated (left)
and comparison (right) samples. 
The horizontal position of each data point indicates the median
SFR of the sources in the bin.
Each SFR sample is further divided into two subsamples, i.e., 
$\mstar$ above (blue points) and below (red points)
the median $\mstar$ of the SFR sample, respectively.
The black lines are the best-fit log-linear model to the
black data points. 
The error bars represent a 1$\sigma$ confidence level.
}
\label{fig:bhar_vs_sfr}
\end{figure*}

We plot the $\bharbar$ as a function of SFR for our bulge-dominated 
and comparison samples, respectively, 
in Fig.~\ref{fig:bhar_vs_sfr} (black points).
In each panel, the bins are chosen to include approximately the
same number of sources, and this approach is to reach similar
$\bharbar$ signal-to-noise (S/N) ratios for the bins.
Adjusting the bins does not change our conclusions qualitatively,
although the statistical scatter of $\bharbar$ measurements
increases.

\begin{figure*}
\includegraphics[width=0.49\linewidth]{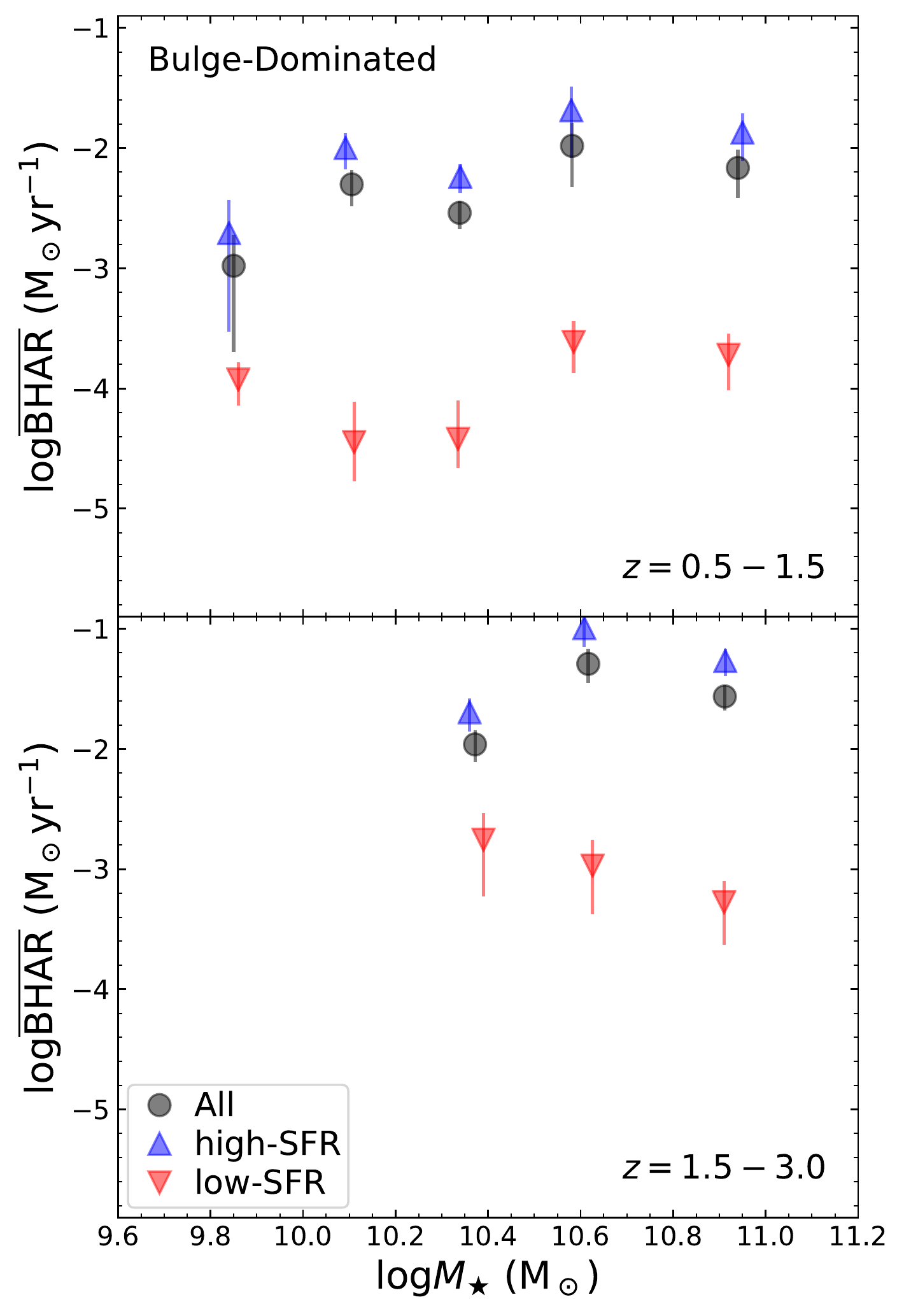}
\includegraphics[width=0.49\linewidth]{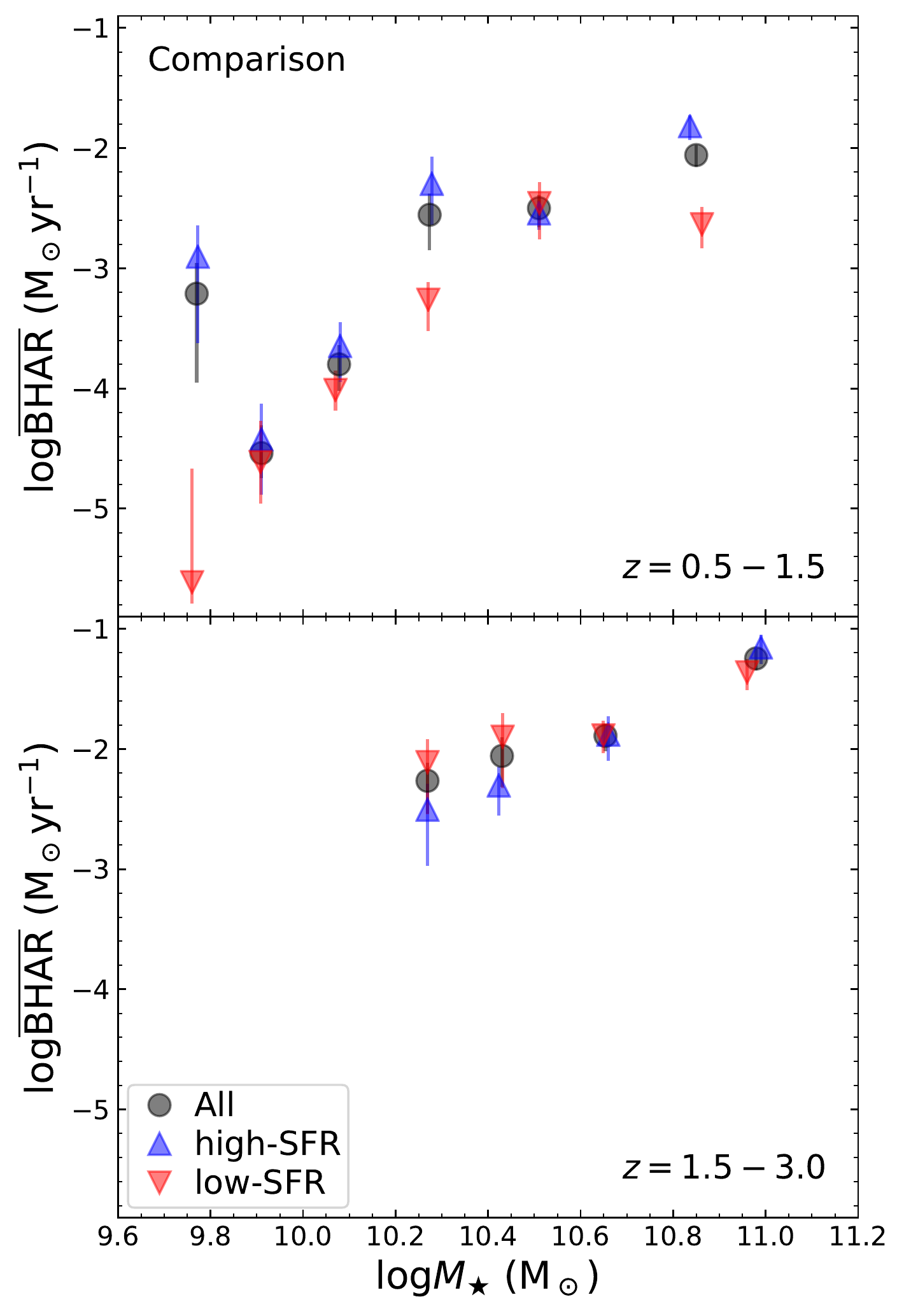}
\caption{Same format as Fig.~\ref{fig:bhar_vs_sfr} but for
$\bharbar$ vs.\ $\mstar$. 
}
\label{fig:bhar_vs_m}
\end{figure*}

For the bulge-dominated sample, $\bharbar$ rises strongly from low 
to high SFR by a factor of $\approx 400$ ($z=0.5\text{--}1.5$) 
and $\approx 100$ ($z=1.5\text{--}3.0$). 
In contrast, for the comparison sample, $\bharbar$ only increases
by a factor of $\approx 10$ ($z=0.5\text{--}1.5$) 
and $\approx 2$ ($z=1.5\text{--}3.0$) from low to high SFR.
We show $\bharbar$ vs.\ $\mstar$ in Fig.~\ref{fig:bhar_vs_m} 
(black points).
For the bulge-dominated sample, there is no strong correlation 
between $\bharbar$ and $\mstar$.
For the comparison sample, $\bharbar$ appears to rise toward 
high $\mstar$ in general.
We note that, due to our limited sample size, statistical 
fluctuations can be strong sometimes.
For example, for the black point at $\log\mstar\approx 9.8$ 
in Fig.~\ref{fig:bhar_vs_m} (right), the $\bharbar$ is mostly 
contributed by a single source.
These fluctuations inevitably cause some scatter in 
Figs.~\ref{fig:bhar_vs_sfr} and \ref{fig:bhar_vs_m}.

\subsection{Is $\bharbar$ Mainly Related to SFR or $\mstar$?}
\label{sec:sfr_or_m}
In this Section, we address the question of whether 
$\bharbar$ is mainly related to SFR or $\mstar$ for the bulge-dominated
and comparison samples, respectively. 
The analysis methods here are similar to those in 
\hbox{\citet{yang17}}. 
We compare our results with \citet{yang17} in Appendix~\ref{app:all}.

In Fig.~\ref{fig:bhar_vs_sfr}, we divide each SFR bin into 
two bins with $\mstar$ above and below the median $\mstar$ 
of the SFR bin, respectively.
In general, the high-$\mstar$ and low-$\mstar$ bins have 
similar $\bharbar$ for the bulge-dominated sample. 
However, the high-$\mstar$ bins have significantly higher 
$\bharbar$ than the corresponding low-$\mstar$ bins for 
the comparison sample. 
Similarly, in Fig.~\ref{fig:bhar_vs_m}, we also divide each
$\mstar$ bin into high-SFR and low-SFR bins. 
The high-SFR bins have much higher $\bharbar$ than the 
corresponding low-SFR bins for the bulge-dominated sample.
In contrast, the high-SFR and low-SFR bins have similar 
$\bharbar$ for the comparison sample. 

The results above qualitatively indicate that $\bharbar$ might 
primarily depend on SFR rather than $\mstar$ for the bulge-dominated 
sample and that the situation is the opposite for the comparison sample.
To further test this point, we perform partial-correlation (PCOR) analyses
with {\sc pcor.r} in the {\sc r} statistical package 
\hbox{\citep{kim15}}.
We first bin sources based on both SFR and $\mstar$ and calculate 
$\bharbar$ for each bin, and Fig.~\ref{fig:bhar_on_grids} shows 
the results.
Following Fig.~\ref{fig:bhar_vs_sfr}, the bins for the $x$-axis
($y$-axis) include similar numbers of sources.
Adjusting the bins does not affect our results qualitatively.
We input the median $\log\mstar$, median $\log\mathrm{SFR}$, 
and $\log\bharbar$ in each bin to {\sc pcor.r} and calculate the 
significance levels for the $\bharbar$-$\mstar$ and $\bharbar$-SFR 
relations, respectively.
The PCOR tests are performed with the Pearson and Spearman
statistics, respectively, and the results are summarized in 
Tab.~\ref{tab:pcor}.
These results show that, for the bulge-dominated sample, the 
$\bharbar$-SFR correlation is significant ($>3\sigma$)
while the $\bharbar$-$\mstar$ correlation is not ($<3\sigma$).
For the comparison sample, the $\bharbar$-$\mstar$ correlation is
significant while the $\bharbar$-SFR correlation is not.
These conclusions are also supported by 
Figs.~\ref{fig:bhar_vs_sfr} and \ref{fig:bhar_vs_m}.
{We note that the lack of a significant 
${\bharbar}$-SFR relation for the comparison sample 
is unlikely to be caused by \xray\ obscuration effects, because the 
effects of obscuration on our $\bharbar$ measurements are 
generally small (see \S\ref{sec:bhar}).
}

Fig.~\ref{fig:bhar_vs_sfr} (left) is the key plot in this paper. 
It displays the strong $\bharbar$-SFR connection and qualitatively 
demonstrates that the $\bharbar$-SFR relation cannot be 
significantly split by $\mstar$. 
We have also tested dividing each SFR bin by other galaxy properties
(instead of $\mstar$) such as $f_{\rm disk}$ (\S\ref{sec:morph}) 
and rest-frame $U-V$ color, and none of these parameters can 
significantly split the $\bharbar$-SFR relation.
Therefore, the strong $\bharbar$-SFR correlation is likely fundamental.

\begin{figure*}
\includegraphics[width=0.49\linewidth]{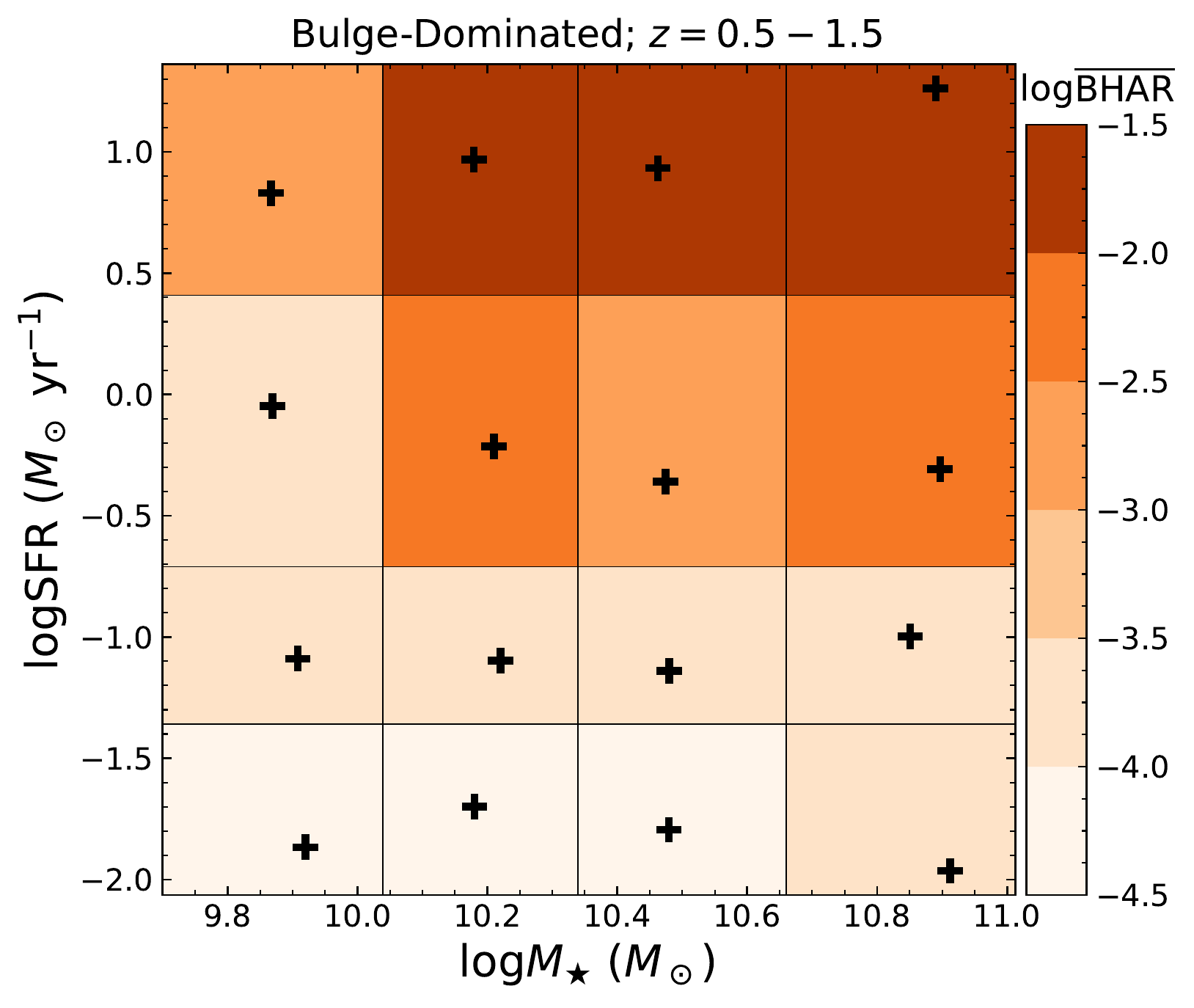}
\includegraphics[width=0.49\linewidth]{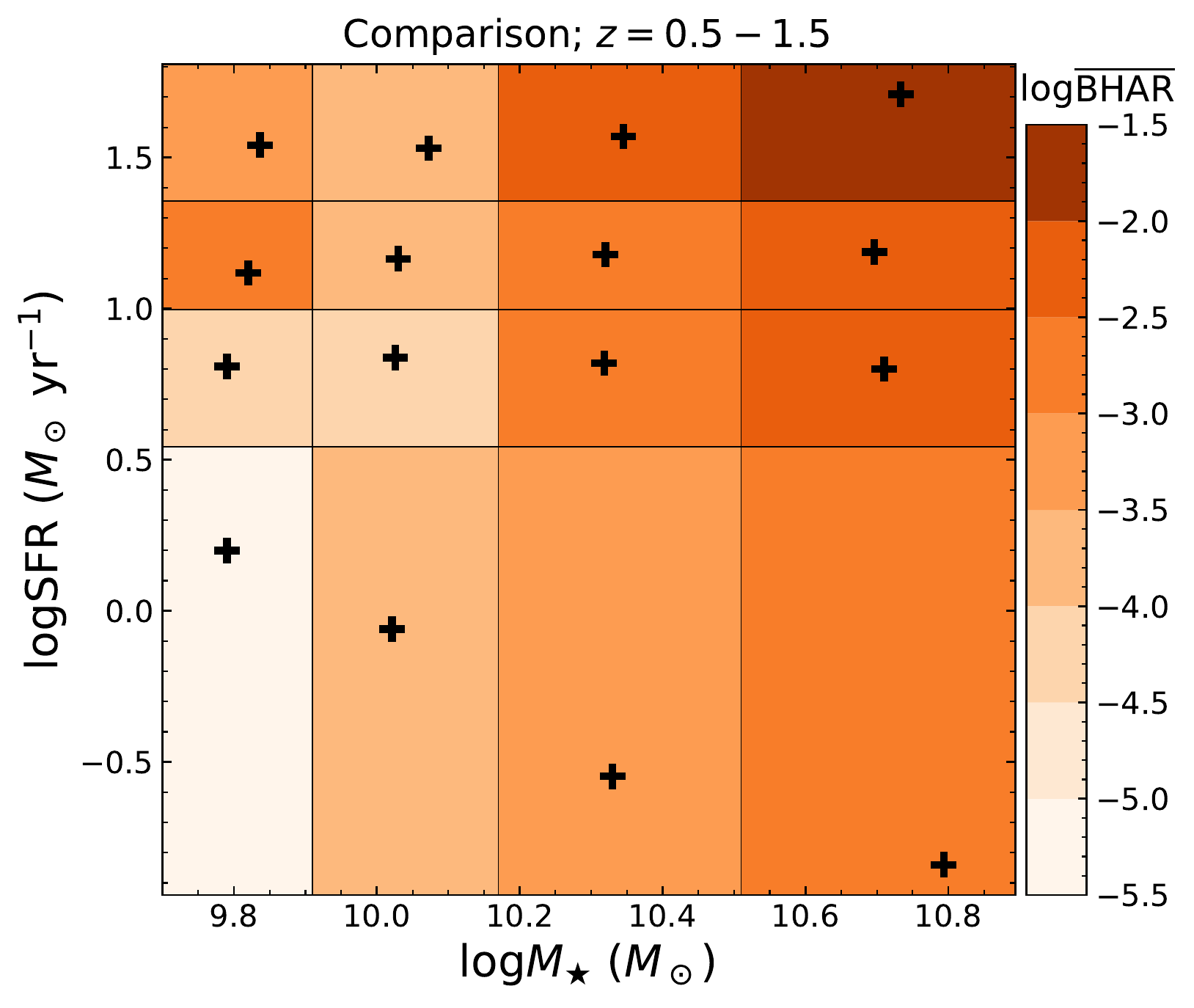}
\includegraphics[width=0.49\linewidth]{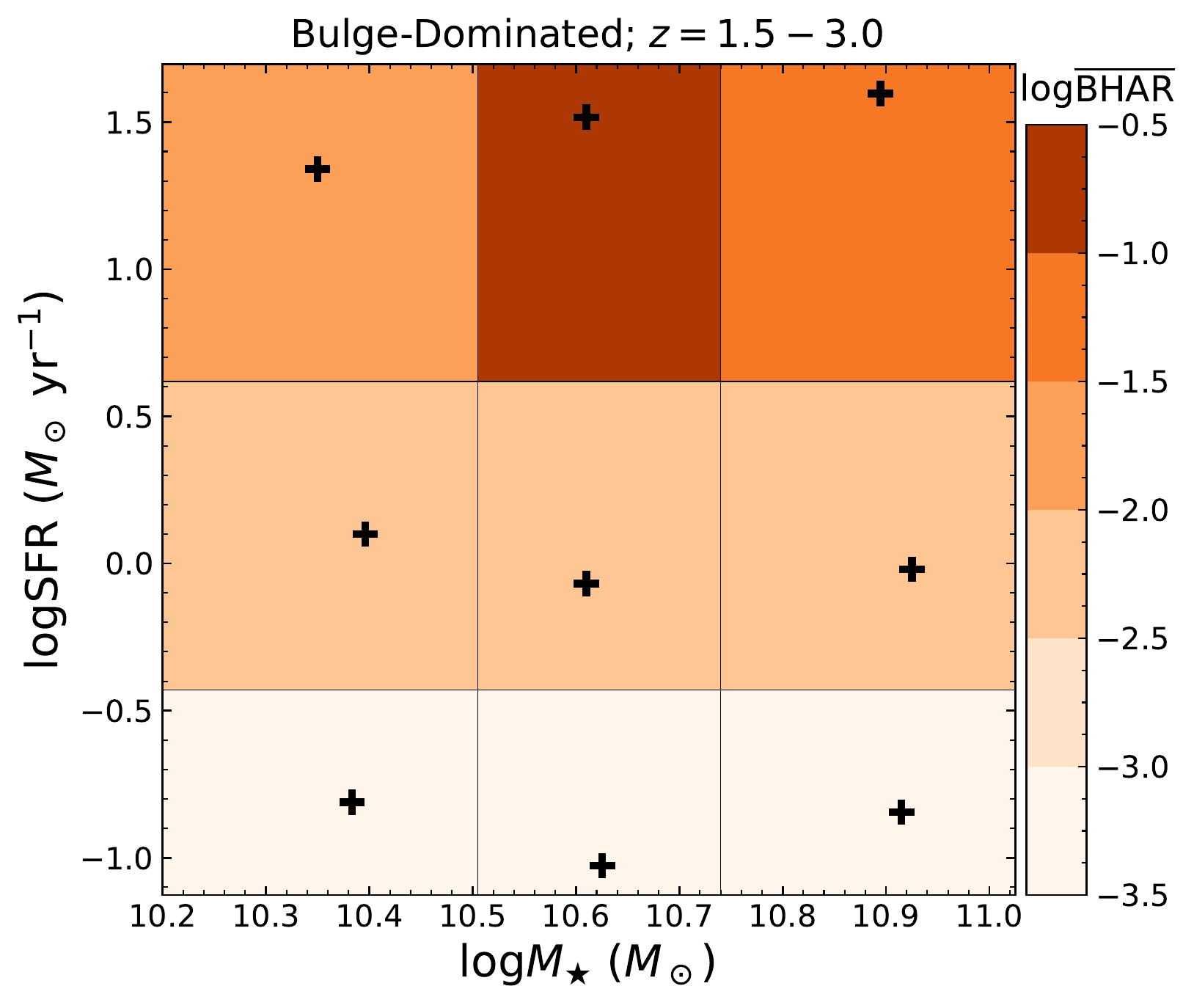}
\includegraphics[width=0.49\linewidth]{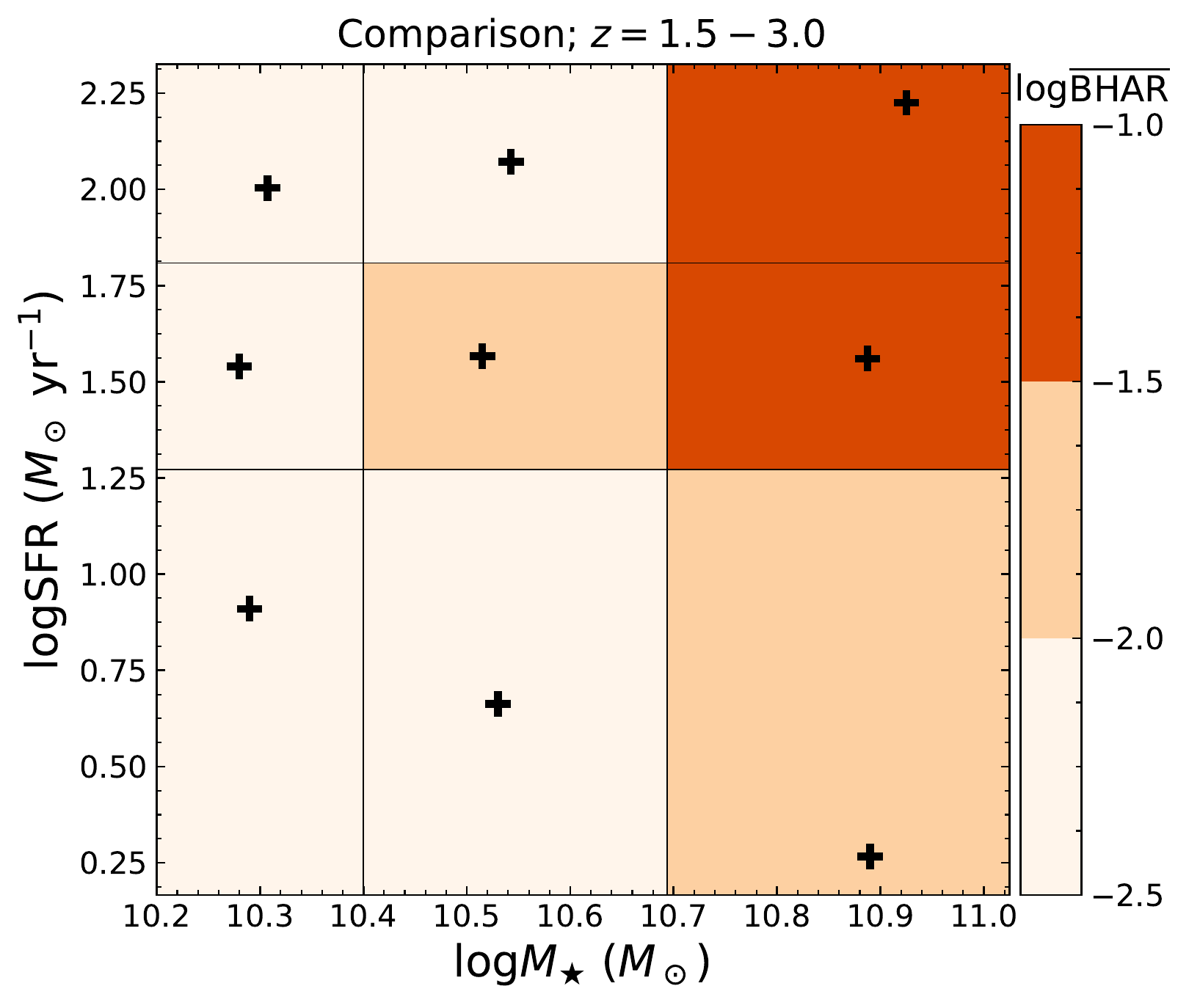}
\caption{Color-coded $\bharbar$ at different $\mstar$ and SFR for 
bulge-dominated (left) and comparison (right) samples. 
The black plus sign indicates the median SFR and $\mstar$ of the 
sources in each bin. 
The $\bharbar$, median $\mstar$, and median SFR are the input in our 
PCOR analyses (\S\ref{sec:sfr_or_m}).
}
\label{fig:bhar_on_grids}
\end{figure*}

\begin{table}
\begin{center}
\caption{$p$-values (significances) of 
	partial-correlation analyses for the bulge-dominated (top) 
 	and comparison (bottom) samples.}
\label{tab:pcor}
\begin{tabular}{cccc}\hline\hline
\multicolumn{3}{c}{Bulge-Dominated; $z=0.5\text{--}1.5$} \\ \hline
Relation & Pearson & Spearman \\
 $\bharbar$-$\mstar$ & $0.03$ ($2.2\sigma$) & $0.02$ ($2.3\sigma$) \\
 $\bharbar$-$\mathrm{SFR}$ & $10^{-18.8}$ ($9.0\sigma$) & $10^{-13.3}$ ($7.5\sigma$) \\
\hline
\multicolumn{3}{c}{Bulge-Dominated; $z=1.5\text{--}3$} \\ \hline
Relation & Pearson & Spearman \\
 $\bharbar$-$\mstar$ & $0.26$ ($1.1\sigma$) & $0.44$ ($0.8\sigma$) \\
 $\bharbar$-$\mathrm{SFR}$ & $10^{-28.5}$ ($11.2\sigma$) & $10^{-5.2}$ ($4.5\sigma$) \\
\hline\hline
\multicolumn{3}{c}{Comparison; $z=0.5\text{--}1.5$} \\ \hline
Relation & Pearson & Spearman \\
 $\bharbar$-$\mstar$ & $10^{-7.7}$ ($5.6\sigma$) & $10^{-8.4}$ ($5.9\sigma$) \\
 $\bharbar$-$\mathrm{SFR}$ & $0.01$ ($2.5\sigma$) & $0.02$ ($2.3\sigma$) \\
\hline
\multicolumn{3}{c}{Comparison; $z=1.5\text{--}3$} \\ \hline
Relation & Pearson & Spearman \\
$\bharbar$-$\mstar$ & $10^{-14.3}$ ($7.8\sigma$) & $10^{-4.5}$ ($4.2\sigma$) \\
 $\bharbar$-$\mathrm{SFR}$ & $0.20$ ($1.3\sigma$) & $0.97$ ($0.0\sigma$) \\
    \hline
\end{tabular}
\end{center}
\end{table}

\subsection{Quantification of the $\bharbar$-SFR Relation}
\label{sec:quan}
In \S\ref{sec:sfr_or_m}, we find that $\bharbar$ is mainly 
correlated with SFR rather than $\mstar$ for the bulge-dominated sample.
To quantify this $\bharbar$-SFR relation, we fit the data points in 
Fig.~\ref{fig:bhar_vs_sfr} (left; black points) with a log-linear model. 
For convenience, we list the sample properties of each data point in 
Tab.~\ref{tab:bin_prop}.
We adopt a standard least-$\chi^2$ fitting method implemented by 
a {\sc python} package {\sc scipy.optimize.curve\_fit}.
We first fit the data points in the two redshift bins independently,
and the results are
\begin{equation}\label{eq:bhar_sfr_sph_flex_z}
\log\bharbar = 
    \begin{cases}
        (0.88\pm 0.07)\log\mathrm{SFR} - (2.56\pm 0.08), &z=0.5\text{--}1.5 \\ 
        (0.89\pm 0.08)\log\mathrm{SFR} - (2.38\pm 0.09), &z=1.5\text{--}3
    \end{cases}
\end{equation} 
Considering the best-fit parameters are similar for the two redshift bins,
we fit all the data points in both redshift bins simultaneously. 
The best-fit model is 
\begin{equation}\label{eq:bhar_sfr_sph_flex}
\begin{split}
    \log\bharbar = (0.92\pm 0.04)\log\mathrm{SFR} - (2.47\pm 0.05),
\end{split}
\end{equation} 
where the errors are calculated under a 68\% confidence level. 
The reduced $\chi^2$ of the fit is 0.8 ($p\text{-value}=53\%$),
showing that the fit quality is acceptable. 
Considering that the slope of the best-fit model is close to unity,
we also fit the data with slope fixed to unity. 
This procedure results in 
\begin{equation}\label{eq:bhar_sfr_sph}
\begin{split}
    \log\bharbar = \log\mathrm{SFR} - (2.48\pm 0.05).
\end{split}
\end{equation} 
The fit quality is also acceptable, with reduced $\chi^2$ of 
1.2 ($p\text{-value}=32\%$).
This best-fit model is displayed in Fig.~\ref{fig:bhar_vs_sfr} 
(left).
The best-fit $\bharbar$/SFR ratio in this model is $10^{-2.48}$.

Our $\bharbar$ does not include the accretion from BL AGNs 
(\S\ref{sec:analyses}).
Here, we consider this missed accretion power statistically.
We first construct a non-BL AGN sample with $\lx$ and redshift 
matched with the spectroscopic BL AGN sample (\S\ref{sec:m_sfr}):
for each BL AGN, we randomly select a ``nearby'' non-BL AGN in 
the $\lx$-$z$ plane (within $\log\lx\pm 0.15$ and $z\pm 0.2$).
We find that $\approx 35\%$ of these non-BL AGNs reside in 
bulge-dominated galaxies. 
Assuming, following the unified model, 
that a similar fraction of BL AGNs have bulge-dominated
hosts, we find that the missed accretion power (contributed by 
BL AGNs) is $\approx 40\%$ of the observed accretion power for
bulge-dominated galaxies.
Therefore, after including the accretion power of BL AGNs, the 
$\bharbar$/SFR ratio might be slightly higher ($\approx 0.15$~dex) 
than the best-fit value.
We note that errors resulting from radiative-efficiency and IMF 
uncertainties likely exist, and thus the best-fit value of $\bharbar$/SFR 
inevitably suffers from systematic uncertainty up to a factor of a few.
However, the systematic uncertainties should not affect our main qualitative 
conclusion, i.e.\ $\bharbar$ primarily depends on SFR among 
bulge-dominated galaxies.

\begin{figure}
\includegraphics[width=\linewidth]{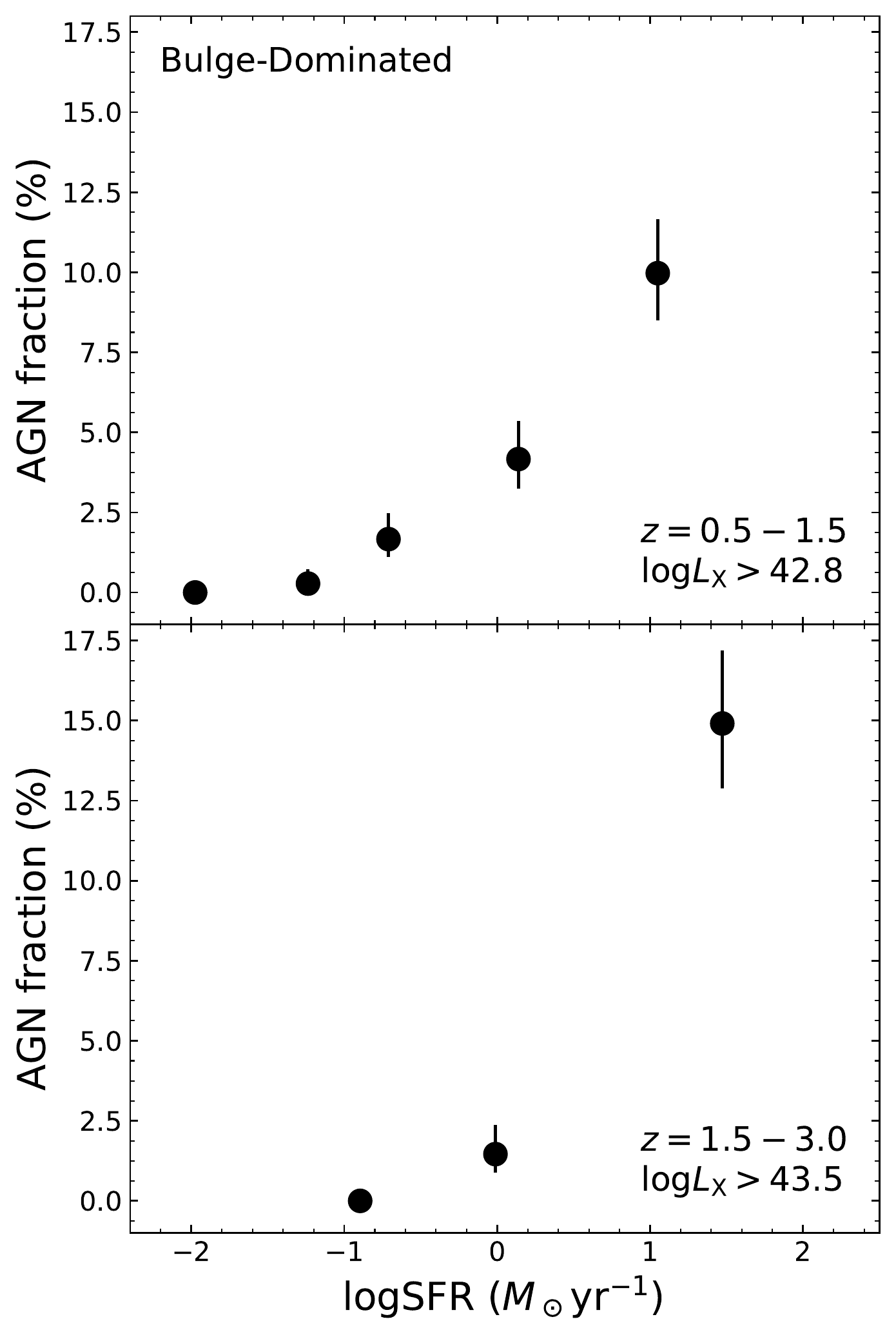}
\caption{AGN fraction as a function of SFR for bulge-dominated galaxies.
The upper and lower panels are for $z=0.5\text{--}1.5$ and $z=1.5\text{--}3$,
respectively. 
As labelled, the AGN fractions are calculated based on different $\lx$ 
thresholds for different redshift bins, and the thresholds are derived 
from the \xray\ flux limit of the COSMOS survey (\S\ref{sec:quan}).
The AGN fraction rises toward high SFR.
}
\label{fig:AGNfrac_vs_sfr_sph}
\end{figure}

Considering the importance of SMBH-galaxy growth among 
bulge-dominated galaxies, we also plot AGN fraction as a function 
of SFR in Fig.~\ref{fig:AGNfrac_vs_sfr_sph}. 
Here, we count an \xray\ source as an ``AGN'' if it has 
$\log\lx > 42.8$ ($z=0.5\text{--}1.5$) or $\log\lx > 43.5$ 
($z=1.5\text{--}3$). 
These thresholds are the $\lx$ limits at $z=1.5$ and $z=3$, 
respectively, for a $0.5\text{--}10$~keV flux limit of 
$8.9\times 10^{-16}$~erg~cm$^{-2}$~s$^{-1}$. 
This flux limit is the detection limit of the COSMOS survey 
\citep{civano16}, which is the shallowest CANDELS \xray\ survey 
(Tab.~\ref{tab:sample}).  
$\bharbar$ is mainly driven by duty cycle and average accretion 
rate of AGNs. 
From Fig.~\ref{fig:AGNfrac_vs_sfr_sph},\footnote{{We cannot 
derive reliable average AGN \xray\ luminosities and thereby 
accretion rates due to the small AGN sample sizes in most bins 
(see Tab.~\ref{tab:bin_prop}).}} 
AGN fraction increases toward high SFR for both redshift bins.
This result indicates that the positive $\bharbar$-SFR relation 
is, at least partially, due to the rise of AGN duty cycle toward 
high SFR.
Detailed quantitative analyses of AGN duty cycle and average 
accretion rate require the full distribution of BHAR (see 
\S\ref{sec:bhar}), for which we leave to future studies.

\begin{table}
\begin{center}
\caption{Properties of each bin in Fig.~\ref{fig:bhar_vs_sfr} (left; black points).}
\label{tab:bin_prop}
\begin{tabular}{cccccc}\hline\hline
\multicolumn{5}{c}{Bulge-Dominated; $z=0.5\text{--}1.5$} \\ \hline
$\log\rm{SFR}$ & $\log\mstar$ & $\log\bharbar$ & Gal.\ \# & X.\ \# & AGN \# \\
(1) & (2) & (3) & (4) & (5) & (6) \\ \hline
$-1.97$ & 10.24 &   $-4.35^{+0.17}_{-0.17}$ &   360 & 14 & 0 \\
$-1.24$ & 10.44 &   $-3.67^{+0.14}_{-0.16}$ &   360 & 20 & 1 \\
$-0.71$ & 10.48 &   $-2.82^{+0.20}_{-0.39}$ &   360 & 29 & 6 \\
$0.14$ & 10.25 &   $-2.41^{+0.12}_{-0.15}$ &   360 & 46 & 15 \\
$1.05$ & 10.19 &   $-1.69^{+0.12}_{-0.17}$ &   361 & 69 & 36 \\
\hline\hline
\multicolumn{5}{c}{Bulge-Dominated; $z=1.5\text{--}3$} \\ \hline
$\log\rm{SFR}$ & $\log\mstar$ & $\log\bharbar$ & Gal.\ \# & X.\ \# & AGN \# \\
(1) & (2) & (3) & (4) & (5) & (6) \\ \hline
$-0.90$ & 10.62 &   $-3.27^{+0.15}_{-0.21}$ &   274 & 7 & 0 \\
$-0.01$ & 10.64 &   $-2.28^{+0.13}_{-0.18}$ &   274 & 21 & 4 \\
$1.47$ & 10.59 &   $-1.08^{+0.09}_{-0.11}$ &   275 & 77 & 41 \\
\hline
\end{tabular}
\end{center}
\begin{flushleft}
{\sc Note.} ---
(1) \& (2) Median $\log\rm{SFR}$ and $\log\mstar$.
(3) $\log\bharbar$ and its uncertainties.
(4) Number of galaxies.
(5) Number of X-ray sources.
(6) Number of AGNs as defined in Fig.~\ref{fig:AGNfrac_vs_sfr_sph}.
\end{flushleft}
\end{table}

\subsection{Reliability Checks}
\label{sec:reli}
Our $\mstar$ and SFR measurements are mostly based on SED fitting 
of rest-frame UV-to-NIR photometry, and we have removed BL AGNs 
from our sample to avoid strong AGN SED components that might affect 
our results (\S\ref{sec:m_sfr} and \S\ref{sec:quan}).
Also, from Fig.~\ref{fig:thumb}, the \xray\ detected sources do
not appear to have strong central point-like emission, indicating 
that the presence of AGNs should not significantly affect the $\mstar$, 
SFR, and morphology measurements.  
On the other hand, if potential AGN SED contamination significantly 
biased our analyses, we would expect to reach a similar conclusion for
bulge-dominated and comparison samples, which is not the case 
(\S\ref{sec:sfr_or_m}).
Therefore, we consider that our conclusions are not biased by AGN
SED contamination.
{In \S\ref{sec:m_sfr}, we have also discussed the SFR uncertainties 
of SED-based and FIR-based measurements, and found that our main 
conclusions are unlikely to be affected by those uncertainties.}

Our bulge-dominated galaxies at $z=0.5\text{--}3$ 
are selected utilizing machine-learning morphological 
measurements based on \hst\ $H$-band imaging 
(\S\ref{sec:morph} and \S\ref{sec:samp}).
Morphological measurements at high redshift are challenging
due to effects such as redshifting of photons and image 
degradation (e.g. \hbox{\citealt{conselice14}} and 
references therein).
Detailed assessment of these redshift effects on our 
results requires careful simulations of \hst\ imaging 
and repetition of the machine-learning measurements 
on the simulated data. 
These procedures are beyond the scope of this work.
Here, we qualitatively discuss the robustness of our 
results against such redshift effects.

Due to the redshifting of photons, the same observed-frame
wavelength covers different rest-frame wavelengths 
at different redshifts.
The correction for this redshifting effect is named
the ``morphological $k$-correction''.
From multiwavelength observations of local 
galaxies, \hbox{\citet{taylor_mager07}} found that 
the morphological $k$-correction is weak in the optical/NIR 
wavelength range
($\approx 0.36 \text{--} 0.85$~$\mu$m, where $0.36\ \mu$m 
corresponds to the Balmer break and $0.85\ \mu$m is the longest 
wavelength available in their work), especially for 
elliptical/S0 galaxies.
For our work, the observed-frame $H$~band (used for
morphological measurements; \S\ref{sec:morph}) does 
not reach out to rest-frame UV photons below the Balmer break 
at $z=0.5\text{--}3$, 
and it corresponds to the rest-frame NIR ($\approx 1\ \mu$m)  
and optical ($\approx 0.4\ \mu$m) light at $z=0.5$ and 
$z=3$, respectively. 
Also, considering that we only utilize the morphological 
information for a basic selection of bulge-dominated 
galaxies rather than, e.g. a quantitative measurement 
of galaxy size, we conclude that the morphological 
$k$-correction should not affect our results 
qualitatively.
Another point of support for this conclusion is that, 
although the $H$ band is sampling different wavelengths for
$z=0.5\text{--}1.5$ and $z=1.5\text{--}3$, we have 
obtained qualitatively the same results for the two redshift
bins (\S\ref{sec:sfr_or_m}).

At low redshift, the $H$~band samples rest-frame red optical/NIR 
photons. 
Since galactic-disk components are generally bluer than bulge 
components, one might worry that $H$-band imaging could miss 
disk components. 
This issue mainly happens for low-$\mstar$ faint disky galaxies.
Considering that our main focus is relatively massive galaxies 
(\S\ref{sec:samp}) and that CANDELS $H$-band data are deep, this
issue might not be problematic for our study.
However, we still check this issue in the following ways.
First, we visually check \hst\ $I$-band 
cutouts of $\approx 100$ random galaxies in our bulge-dominated 
samples at $z=0.5\text{--}1$.
We do not find significant disk components for these sources.
Indeed, the average $I$-band S\'ersic index of our low-redshift 
bulge-dominated sample is $\approx 3.5$ (using the measurements 
of \citealt{scarlata07}), 
which is typical for bulge-dominated galaxies 
\citep[e.g.][]{buitrago13, conselice14}.
On the other hand, we visually check the low-redshift disky 
($f_{\rm disk} \geq 2/3$; \S\ref{sec:samp}) galaxies in our sample. 
We find their disk components do not appear to be significantly 
weaker in $H$ band than in $I$ band, and we attribute this result
to the deep exposure and relatively low extinction of $H$ band.
Therefore, we consider that the $H$ band-based morphological 
classification is robust at low redshift.

The imaging quality generally becomes worse toward higher 
redshift.
Some fine galactic structures (e.g. spiral 
arms and clumps) might be smoothed out, and thus some disky
and irregular galaxies might be classified as the smooth 
bulge-dominated type (e.g. \hbox{\citealt{mortlock13}}).
Therefore, our bulge-dominated sample might be 
``contaminated''.
However, this issue is, at least to some extent, 
mitigated by the $H$~mag cut ($H<24.5$) applied to our 
sample (\S\ref{sec:morph}).
This cut guarantees a minimum signal-to-noise ratio
(S/N~$\approx 80$) of the imaging, with the penalty of a 
smaller sample size. 
Also, if our bulge-dominated sample were strongly 
contaminated, we would observe a similar
$\bharbar$-SFR-$\mstar$ relation for both 
the bulge-dominated and comparison samples. 
However, the $\bharbar$ dependences on SFR and $\mstar$
are qualitatively different for these two samples 
(\S\ref{sec:sfr_or_m}).
We thus consider that image degradation should not 
be a significant issue for our conclusions.

\section{Discussion}
\label{sec:discuss}
\subsection{Physical Implications}
\label{sec:phy}
We emphasize that the $\bharbar$-SFR correlation only exists
for our bulge-dominated sample, while $\bharbar$ appears to be primarily
correlated with $\mstar$ for the comparison sample.
This difference indicates that SMBHs only coevolve with bulges rather
than entire galaxies, consistent with the observations of local
systems \citep[e.g.][]{kormendy13, davis18}.
Such SMBH-bulge coevolution might be driven by the amount of cold 
gas available in the bulge, since both SMBH and bulge growth require
cold gas.
From the SMBH-bulge coevolution scenario, we expect that
$\bharbar$ is also fundamentally correlated with bulge SFR
even when a galactic disk is present.

Earlier studies speculated an intrinsic 
$\bharbar$-SFR relation for the overall galaxy population 
(see \S\ref{sec:intro}).
However, the scenario of SMBH vs.\ entire galaxy coevolution
leads to $\mbh$ being strongly related to $\mstar$ rather than
$\mbulge$ in the local universe, contradicting
observations \citep[e.g.][]{kormendy13}.
To reconcile this contradiction, an \textit{ad hoc} galaxy 
evolution model was invoked where all stellar mass formed
in the distant universe ($z\gtrsim 0.5$) is transformed to 
bulge mass at $z=0$ \citep{jahnke09, mullaney12}.
In contrast, the $\bharbar$-bulge SFR correlation, as revealed
by our work, can naturally result in the $\mbh$-$\mbulge$
relation observed in the local universe, without invoking any
unphysical galaxy evolution models.
Our findings highlight the critical role of morphological 
measurements when observationally studying the connections 
between distant SMBHs and their host galaxies, as the $\bharbar$-SFR
correlation only exists among bulge-dominated galaxies.
Without deep \hst\ observations of CANDELS, our discovery 
would not be possible (see Appendix~\ref{app:all}).

Some papers attribute the local $\mbh$-$\mbulge$ relation entirely 
to a non-causal statistical origin (e.g. \citealt{peng07}; 
\citealt{jahnke11}). 
If galaxy/SMBH mergers happen frequently enough, the scatter of the
$\mbh$-$\mbulge$ relation could be averaged out. 
Our results show that there is indeed an intrinsic $\bharbar$-SFR
connection at high redshift that can lead to the $\mbh$-$\mbulge$ 
correlation among nearby galaxies (\S\ref{sec:local}).
Therefore, the non-causal scenarios of merger averaging are not 
necessary to explain the $\mbh$-$\mbulge$ relation. 
Also, recent observations of \cite{yang18} show that frequent 
mergers will lead to a $\mbh/\mbulge$ ratio much smaller than 
the observed values in the local universe.

{
\hbox{\citet{kocevski17}} found that, for compact galaxies, 
the star-forming population has elevated AGN fraction compared 
to the quiescent population with matched ${\mstar}$ at 
${z\approx 2}$. 
However, for extended galaxies, the star-forming and quiescent 
populations have similar AGN fractions. 
Since our bulge-dominated population is morphologically more 
compact than other galaxy populations in general 
(see Fig.~\ref{fig:thumb}; e.g. \citealt{huertas_company15}),
our results in Fig.~\ref{fig:bhar_vs_m} are broadly consistent 
with the findings of \hbox{\citet{kocevski17}}.
While we consider our results as evidence of SMBH-bulge
coevolution, \cite{kocevski17} argued 
that a contraction process might trigger both compact 
starburst activity and SMBH accretion. 
In our scenario, bulge SFR is fundamentally correlated with 
${\bharbar}$;
in their scenario, compactness is a critical galaxy property 
linked with SMBH growth.
To address the question of which scenario is more physical, 
one needs to break the degeneracy that bulge-dominated 
systems are generally compact.
We will perform these analyses in a future paper 
(Ni et al.\ in prep.).
}

\subsection{Implications for the $\mbh$-$\mbulge$ Relation}
\label{sec:local}
From the best-fit results in \S\ref{sec:quan}, we have
$\bharbar/\mathrm{SFR} = 10^{-2.48}$.
This value is similar to the typical observed $\mbh/\mbulge$ values
in the local universe ($\approx 10^{-2.5}\text{--}10^{-2.2}$; 
\citealt{kormendy13}).
Also, similar to the observed $\mbh$-$\mbulge$ relation in the local
universe, our $\bharbar$-SFR relation for bulge-dominated galaxies
has slope close to unity.
These similarities indicate that the observed $\mbh$-$\mbulge$ relation
originates from SMBH-bulge coevolution as revealed by our work, and
the $\mbh$-$\mbulge$ relation is not heavily biased by the possibility
that observations tend to select massive SMBHs for $\mbh$
measurements \citep[e.g.][]{shankar16}.

The strong $\bharbar$-SFR relation among bulge-dominated galaxies 
indicates that SMBH and bulge growth are in lockstep. 
A natural consequence from this lockstep growth is that the 
$\mbh$-$\mbulge$ relation should not have strong redshift dependence.
Some observations suggest that the $\mbh/\mbulge$ ratio appears to be
higher toward higher redshifts \citep[e.g.][]{shields06, ho07}, 
contradicting the scenario of lockstep growth.
However, this apparent redshift dependence of $\mbh/\mbulge$ might
result from observational biases \citep[e.g.][]{lauer07}, because 
$\mbh$ measurements in the distant universe are generally limited 
to luminous quasars.
These luminous quasars are likely the most massive SMBHs accreting 
at high Eddington ratios, and thus the observed $\mbh/\mbulge$ 
should be systematically higher than the typical $\mbh/\mbulge$ among
the entire galaxy population.

\subsection{Galaxies that are Not Bulge-Dominated}
\label{sec:non_sph}
{
For our bulge-dominated sample, ${\bharbar}$ is fundamentally related to SFR. 
In contrast, for our comparison sample consisting of galaxies that are not 
bulge-dominated (\S\ref{sec:samp}), ${\bharbar}$ is not strongly coupled with SFR 
(\S\ref{sec:sfr_or_m}), likely due to the fact that their total SFR is mostly 
contributed by non-bulge components.
Actually, most (${\approx 80\%}$) of the comparison galaxies are 
irregular/disk-dominated galaxies with no significant bulge components 
(${f_{\rm sph}<2/3}$; \S\ref{sec:samp}).
The rest (${\approx 20\%}$) of the population in the comparison sample 
is bulge-disk systems.
For these systems, according to the SMBH-bulge coevolution scenario, 
${\bharbar}$ should be intrinsically correlated 
with bulge SFR (\S\ref{sec:phy}).
}

{For our comparison sample, ${\bharbar}$ is strongly 
related to ${\mstar}$ (\S\ref{sec:sfr_or_m}).
The implications of this ${\bharbar}$-${\mstar}$ 
relation are discussed in detail by \cite{yang18},\footnote{{Although \cite{yang18} 
focused on the ${\bharbar}$-${\mstar}$
relation for the entire galaxy population, their conclusions should largely hold
for our comparison galaxies which are numerically the main population 
(${\approx 75\%}$; \S\ref{sec:samp} and Appendix~\ref{app:all}).}} 
and we only summarized their main points below. 
\cite{yang18} found that ${\bharbar}$/${\sfrbar}$ rises 
toward high ${\mstar}$, i.e.
massive galaxies are more effective in feeding their SMBHs (see their \S4.2).
This result inevitably leads to a higher ${\mbh/\mstar}$ ratio for more massive 
galaxies in the local universe, i.e. the typical local ${\mbh}$-${\mstar}$ relation 
should be non-linear (see their \S4.3 and \S4.4).
On the other hand, \cite{yang18} also considered that the local ${\mbh}$-${\mstar}$ 
relation might not be tight due to different stellar-mass histories of local 
galaxies with similar ${\mstar}$ (see their \S3.4.1). 
This is because ${\bharbar}$ is higher toward high redshift, at a 
given ${\mstar}$ (see their Fig.~9). 
Therefore, for two galaxies with similar ${\mstar}$ in the local universe, the 
one that forms at higher redshift should have a more massive SMBH.
}

\section{Summary and Future Work}
\label{sec:sum}
We have studied the $\bharbar$ dependence on SFR and $\mstar$ 
for a bulge-dominated sample and a comparison sample of galaxies, 
respectively,
based on multiwavelength observations of the CANDELS fields.
Our main analysis procedures and conclusions are summarized 
below:

\begin{enumerate}

\item We have compiled redshift, $\mstar$, and SFR for galaxies 
brighter than $H=24.5$ from the CANDELS catalogs (\S\ref{sec:m_sfr}). 
The CANDELS $\mstar$ and SFR measurements are based on SED fitting.
For sources detected by \herschel, we estimate their SFR from 
FIR photometry.
We have applied $\mstar$ cuts of $\log\mstar > 9.7$ ($z=0.5\text{--}1.5$) 
and $\log\mstar > 10.2$ ($z=0.5\text{--}1.5$) to our sample to 
ensure $\mstar$ completeness (\S\ref{sec:samp}).
Based on machine-learning morphological measurements 
(\S\ref{sec:morph}), we have selected a sample of bulge-dominated
galaxies and included the other galaxies in a comparison sample
(\S\ref{sec:samp}). 
The bulge-dominated galaxies consist of $\approx 25\%$ of the 
entire galaxy population.

\item We have measured sample-averaged BHAR for different samples
of galaxies based on the deep \xray\ observations from \chandra\ 
(\S\ref{sec:bhar}).
We first measure the $\lx$ for each \xray\ detected source as 
well as average \xray\ luminosity for undetected sources 
via a stacking process. 
From these measurements, we calculate average AGN bolometric 
luminosity adopting an $\lx$-dependent bolometric correction. 
Finally, we estimate $\bharbar$ from $\overline{\lbol}$ assuming 
a constant radiation efficiency.

\item For the bulge-dominated sample, we have shown, with both
qualitative and quantitative (PCOR) analyses, that $\bharbar$ 
primarily depends on SFR rather than $\mstar$ 
(\S\ref{sec:bhar_vs_sfr_m} and \S\ref{sec:sfr_or_m}). 
For the comparison sample, the situation is the opposite. 
The tight $\bharbar$-SFR connection for bulge-dominated galaxies 
indicates that SMBHs only coevolve with bulges rather than entire
host galaxies (S\S\ref{sec:phy}).
The non-causal scenarios of merger averaging are unlikely the 
origin of the $\mbh$-$\mbulge$ relation in the local universe.

\item Our best-fit $\bharbar$-SFR relation for the bulge-dominated
sample is $\log\bharbar = \log\mathrm{SFR} - (2.48\pm0.05)$ where
the slope is fixed to unity (\S\ref{sec:quan}). 
Our best-fit $\bharbar$/SFR ratio is similar to the observed 
$\mbh/\mbulge$ ratio in the local universe (\S\ref{sec:local}).
This agreement indicates that our observed $\bharbar$-SFR relation 
is indeed responsible for the well-known tight $\mbh$-$\mbulge$ 
correlation among local galaxies.
On the other hand, our findings support that the observed 
local $\mbh$-$\mbulge$ relation is not heavily biased.
The strong $\bharbar$-SFR relation among bulge-dominated galaxies
indicate lockstep growth of SMBHs and bulges, predicting that 
the $\mbh$-$\mbulge$ relation should not have strong redshift 
dependence.

\end{enumerate}

{This paper probes the redshift range of ${z=0.5\text{--}3}$. 
Future studies can extend our work down to ${z\approx 0.2}$
using the 2~deg$^2$ COSMOS field, or even to the local universe 
\citep[e.g.][]{goulding17} based on wide surveys, e.g. 
\hbox{XMM-XXL} \citep{pierre16}, 
Stripe~82X \citep{lamassa16},
\hbox{XMM-SERVS} \citep{chen18}, and
the \chandra\ Source Catalog \citep{evans10}.
Compared to distant systems in deep fields, local sources 
have the advantages of larger sample sizes and more accurate 
morphological measurements, and these advantages could reduce the 
uncertainties of the ${\bharbar}$-SFR relation significantly.
In the near future, we will also investigate whether bulge SFR or 
galaxy compactness is more tightly linked to SMBH growth 
(Ni et al. in prep.; \S\ref{sec:phy}).
Future work could furthermore derive the full BHAR distribution as a function
of SFR and ${\mstar}$ for bulge-dominated and comparison galaxies,
respectively, and detailed sample properties such as 
duty cycle and average accretion rate of AGNs can be further obtained
and analyzed (\S\ref{sec:bhar} and Fig.~\ref{fig:AGNfrac_vs_sfr_sph}).
However, such studies will require a large galaxy sample with reliable
morphological measurements, and thus deep \hst\ (or future \jwst\ and
\textit{WFIRST}) imaging over much larger fields than CANDELS is 
needed.
} 

{Since our results indicate that SMBHs grow in lockstep with 
host-galaxy bulges, we also expect a strong connection between 
${\bharbar}$ and bulge SFR for systems that have both bulge and 
disk components (\S\ref{sec:phy} and \S\ref{sec:non_sph}).
Future ALMA observations could study the ${\bharbar}$-bulge 
SFR connection among these systems.
ALMA can cover FIR wavelengths down to observed-frame ${300\ \mu}$m, 
corresponding to the typical SED-peak wavelength 
(${\approx 100\ \mu}$m, 
rest-frame) of cold-dust emission of galaxies at ${z\approx 2}$.
Therefore, IR luminosities and thereby FIR-based SFR can be reliably 
estimated for these systems with ALMA.
Since ALMA can reach \hst-like resolutions, it should be able to separate 
reliably bulge SFR from total SFR.
The strong ${\bharbar}$-SFR connection among bulge-dominated galaxies 
might be physically driven by the amount of cold gas available 
(see \S\ref{sec:phy}). 
To test this idea, one could compare the gas masses of 
high-SFR vs.\ low-SFR bulge-dominated galaxies with observations by ALMA.
ALMA could measure gas masses by observing the CO lines.
ALMA could also observe the Rayleigh–Jeans tail 
of cold dust emission (${\approx 500\ \mu}$m, rest-frame), 
which is a reliable tracer of dust masses \citep[e.g.][]{scoville17}.
Gas masses can then be estimated from dust masses with the assumption of a 
typical dust-to-gas ratio, although uncertainties inevitably exist in this
conversion \citep[e.g.][]{simpson15}.
}


\appendix
\section{Results for All Galaxies}
\label{app:all}
\begin{figure}
\includegraphics[width=\linewidth]{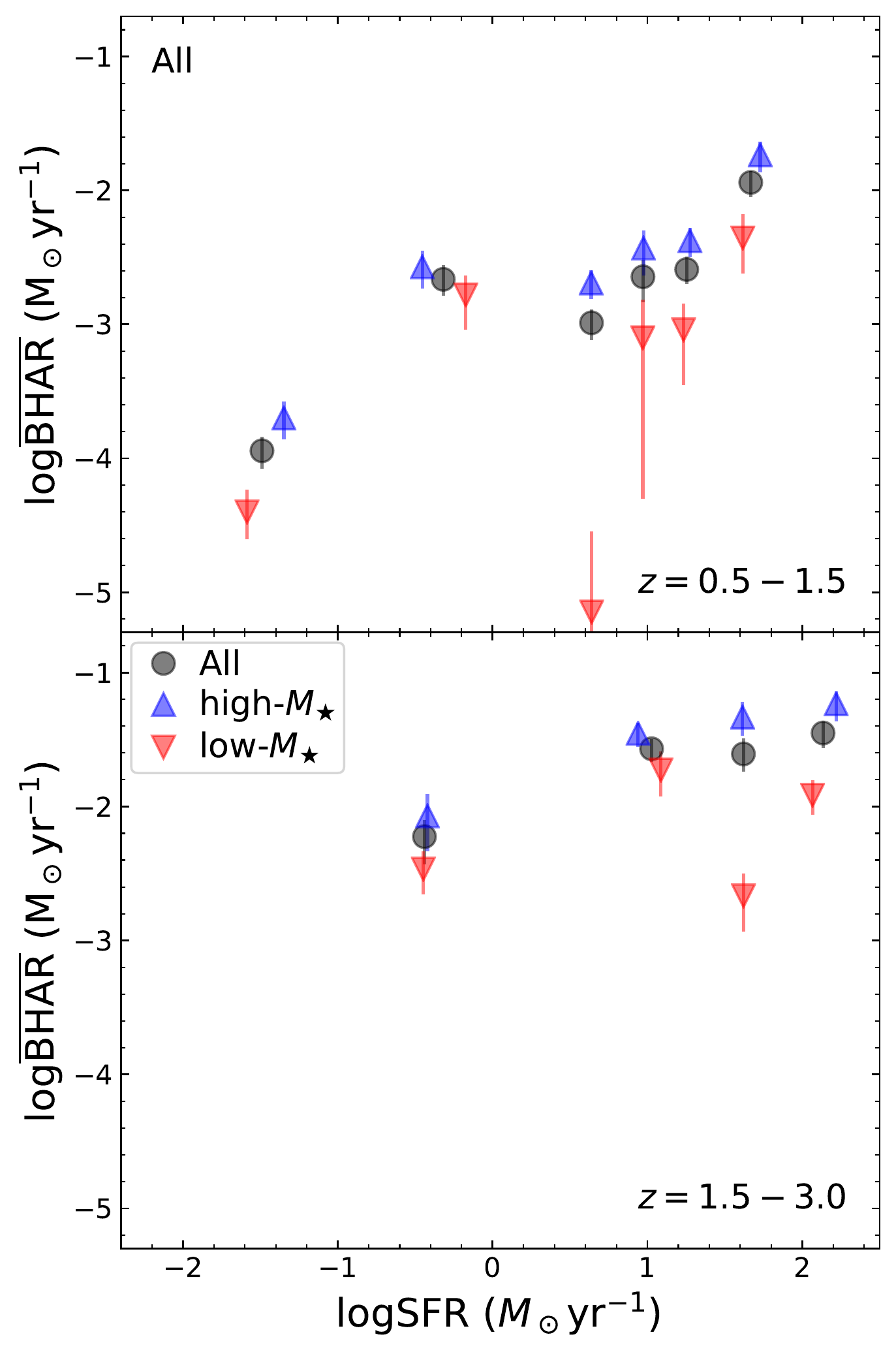}
\caption{Same format as Fig.~\ref{fig:bhar_vs_sfr} but for
all galaxies in our sample. 
}
\label{fig:bhar_vs_sfr_all}
\end{figure}

\begin{figure}
\includegraphics[width=\linewidth]{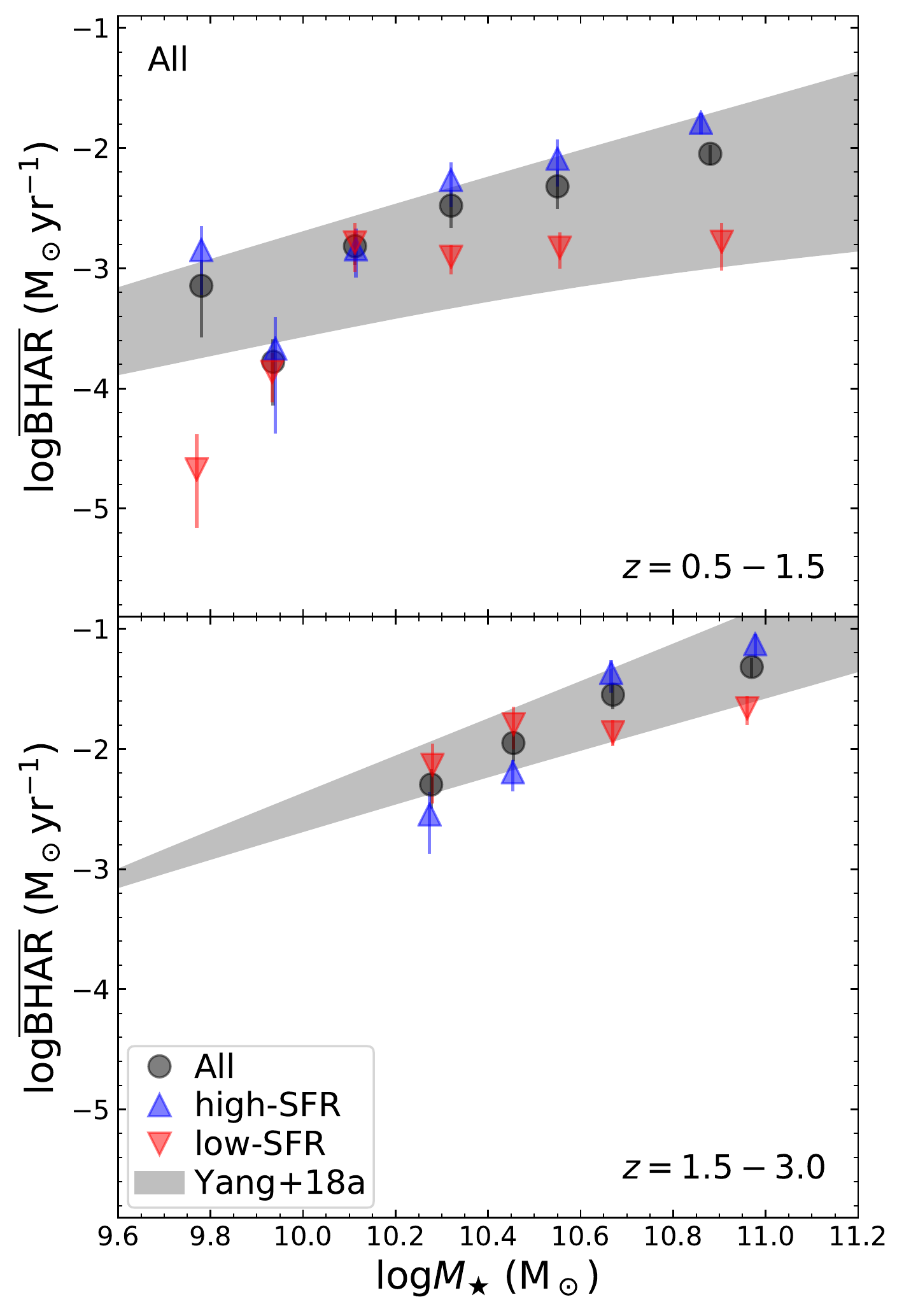}
\caption{Same format as Fig.~\ref{fig:bhar_vs_m} but for
all galaxies in our sample. 
The shaded regions indicate the $\bharbar$-$\mstar$ relation
derived in \hbox{\citet{yang18}}. 
The upper and lower boundaries of the shaded regions indicate 
the $\bharbar$-$\mstar$ relations at the upper and lower 
redshift limits, respectively. 
}
\label{fig:bhar_vs_m_all}
\end{figure}

\begin{figure}
\includegraphics[width=\linewidth]{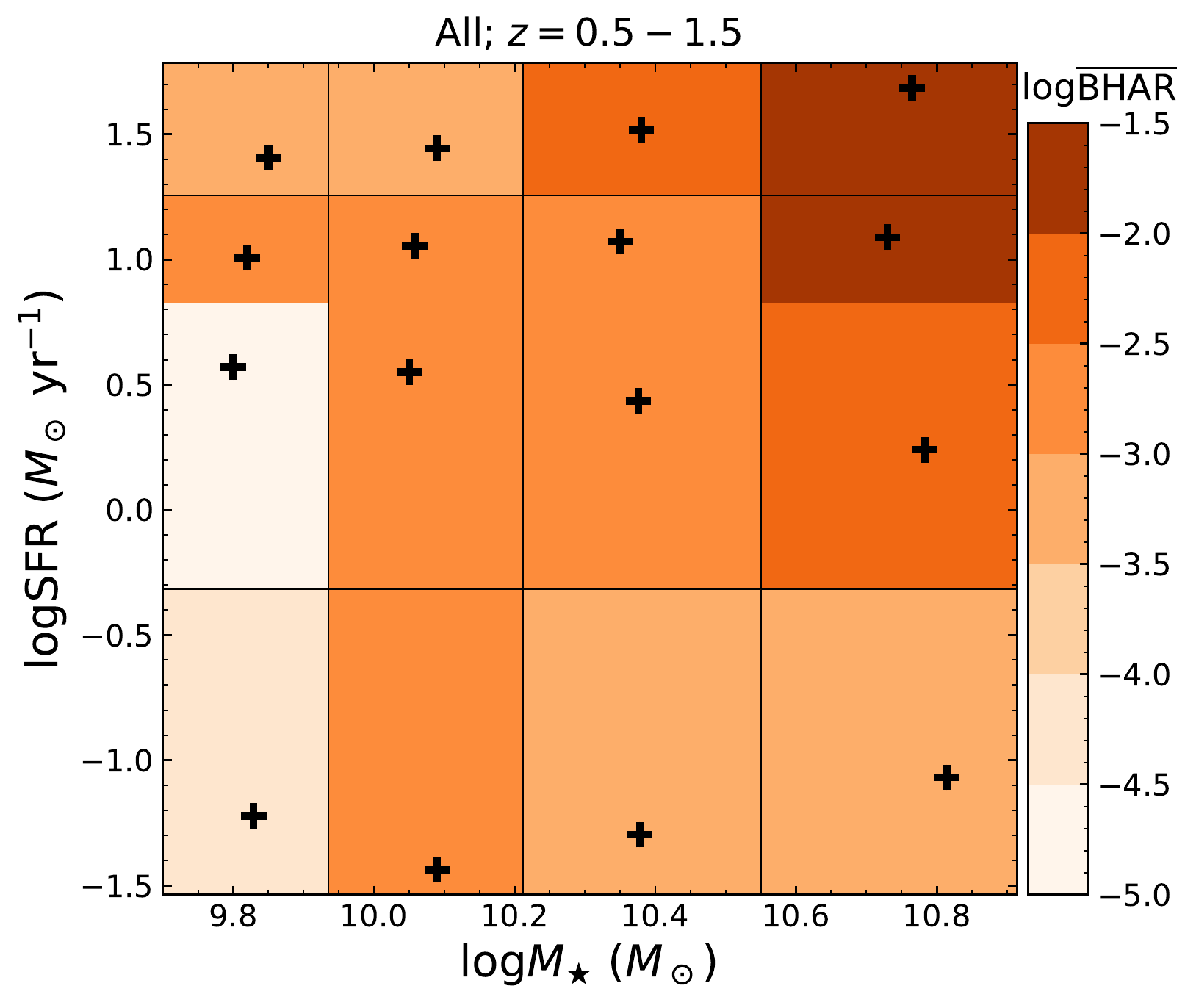}
\includegraphics[width=\linewidth]{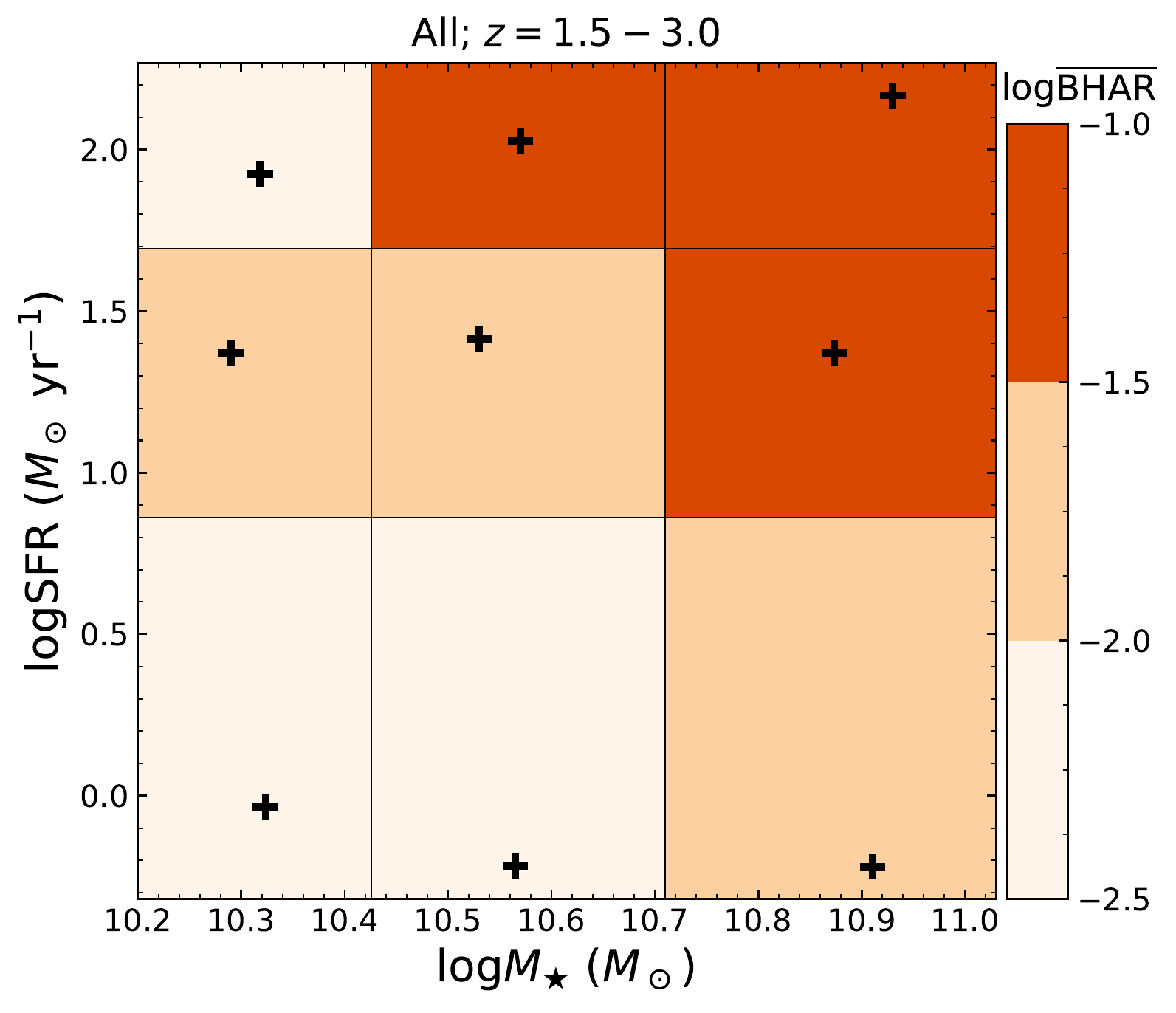}
\caption{Same format as Fig.~\ref{fig:bhar_on_grids} but for
all galaxies in our sample.}
\label{fig:bhar_on_grids_all}
\end{figure}

In this appendix, we perform analyses like those in \S\ref{sec:res} 
for all galaxies with $H<24.5$, including both the bulge-dominated and 
comparison samples grouped together.  
The results are presented in Figs.~\ref{fig:bhar_vs_sfr_all}, 
\ref{fig:bhar_vs_m_all}, and \ref{fig:bhar_on_grids_all}, 
and Tab.~\ref{tab:pcor_all}.
In Fig.~\ref{fig:bhar_vs_m_all}, we also compare the $\bharbar$-$\mstar$
relation with that derived in \hbox{\citet{yang18}}. 
The $\bharbar$-$\mstar$ relation in this work agrees
with the results of \hbox{\citet{yang18}}.

From Tab.~\ref{tab:pcor_all}, $\bharbar$ is more strongly related 
to $\mstar$ than SFR. 
This is expected, because the comparison sample is the numerically 
dominant galaxy population (see \S\ref{sec:samp}) and $\bharbar$ is 
mainly related to $\mstar$ for the comparison sample, especially 
at $z=1.5\text{--}3.0$.
This conclusion is also qualitatively consistent with \hbox{\citet{yang17}}, 
although their statistical significances of the $\bharbar$-$\mstar$ relation 
are higher than those in Tab.~\ref{tab:pcor_all}. 
We attribute this difference to the fact that the dynamic range of 
$\mstar$ probed in \hbox{\citet{yang17}} is much wider than  
that in this work ($\log\mstar \approx 8 \text{--} 11$ vs.\ 
$\log\mstar \approx 10 \text{--} 11$), since here we require $H<24.5$
to ensure high-quality morphological information for all galaxies.
The narrower dynamic range also results in smaller sample
sizes, leading to the relatively large statistical scatter in 
Figs.~\ref{fig:bhar_vs_sfr_all} and \ref{fig:bhar_vs_m_all}
(compared to Figs.~4 and 5 in \hbox{\citealt{yang17}}).

\begin{table}
\begin{center}
\caption{$p$-values (significances) of 
	partial-correlation analyses for all galaxies.}
\label{tab:pcor_all}
\begin{tabular}{ccc}\hline\hline
\multicolumn{3}{c}{All; $z=0.5\text{--}1.5$} \\ \hline
Relation & Pearson & Spearman \\
 $\bharbar$-$\mstar$ & $10^{-5.3}$ ($4.6\sigma$) & $10^{-2.8}$ ($3.1\sigma$) \\
 $\bharbar$-$\mathrm{SFR}$ & $10^{-2.5}$ ($2.9\sigma$) & $0.03$ ($2.2\sigma$) \\
\hline\hline
\multicolumn{3}{c}{All; $z=1.5\text{--}3.0$} \\ \hline
Relation & Pearson & Spearman \\
 $\bharbar$-$\mstar$ & $10^{-4.4}$ ($4.1\sigma$) & $10^{-3.0}$ ($3.3\sigma$) \\
 $\bharbar$-$\mathrm{SFR}$ & $10^{-2.0}$ ($2.6\sigma$) & $0.04$ ($2.1\sigma$) \\
\hline
\end{tabular}
\end{center}
\end{table}


\section*{Acknowledgements}
We thank the referee for helpful feedback that improved this work.
We thank Robin~Ciardullo, Mara~Salvato, Ian Smail, Yongquan~Xue, 
and Wenfei~Yu for helpful discussions.
GY, WNB, and QN acknowledge support from Chandra \xray\ Center
grant AR8-19016X, NASA grant NNX17AF07G, NASA ADP grant 80NSSC18K0878,
and the V.M. Willaman Endowment. The Chandra Guaranteed Time
D.M.A.\ acknowledges the Science and Technology Facilities Council
through grant ST/L00075X/1.
FV acknowledge financial support from CONICYT and CASSACA through the 
Fourth call for tenders of the CAS-CONICYT Fund.
Observations (GTO) for the GOODS-N were selected by the ACIS Instrument
Principal Investigator, Gordon P. Garmire, currently of the Huntingdon
Institute for X-ray Astronomy, LLC, which is under contract to the
Smithsonian Astrophysical Observatory via Contract SV2-82024. 
This project uses Astropy (a Python package; see \citealt{astropy}).




\bibliographystyle{mnras}
\bibliography{all.bib} 

\bsp	
\label{lastpage}
\end{document}